\documentclass[tradiabstract,longauth]{aa}
\usepackage[varg]{txfonts}
\usepackage{graphicx}
\usepackage{lscape}
\usepackage{hyperref}
\usepackage{natbib}
\usepackage{multirow}
\usepackage{color}

\newcommand{\jplus}{\mbox{J-PLUS\,}}
\newcommand{\jpas}{\mbox{J-PAS\,}}

\bibpunct{(}{)}{;}{a}{}{,}
\begin{document}
\author{
A.~J.~Cenarro\inst{1},
M.~Moles\inst{2},
D.~Crist\'obal-Hornillos\inst{1},
A.~Mar\' \i n-Franch\inst{1},
A.~Ederoclite\inst{1},
J.~Varela\inst{1},
C.~L\'opez-Sanjuan\inst{1},
C.~Hern\'andez-Monteagudo\inst{1},
R.~E.~Angulo\inst{2},
H.~V\'azquez Rami\'o\inst{2}
% ...
}
\institute{\small
Centro de Estudios de F\'{\i}sica del Cosmos de Arag\'on (CEFCA) - Unidad Asociada al CSIC, Plaza San Juan, 1, E-44001, Teruel, Spain\goodbreak
\and
Centro de Estudios de F\'{\i}sica del Cosmos de Arag\'on (CEFCA), Plaza San Juan, 1, E-44001, Teruel, Spain\goodbreak
% ...
}

\title{J-PLUS: The Javalambre Photometric Local Universe Survey}
\date{\today}
\abstract
{The {\it Javalambre-Photometric Local Universe Survey} (\jplus) is an ongoing 12-band photometric optical survey, observing thousands of square degrees of the Northern hemisphere from the dedicated JAST/T80 telescope at the {\it Observatorio Astrof\'\i sico de Javalambre} (OAJ). T80Cam is a 2\,deg$^2$ field-of-view camera mounted on this 83\,cm-diameter telescope, and is equipped with a unique system of filters spanning the entire optical range ($3\,500$--$10\,000$\,\AA). This filter system  is a combination of  broad, medium and narrow-band filters, optimally designed to  extract the rest-frame spectral features (the 3\,700--4\,000\,\AA\, Balmer break region, H$\delta$, Ca H+K, the G-band, the Mg$b$ and Ca triplets) that are key to both characterize stellar types and to deliver a low-resolution photo-spectrum for each pixel of the sky observed. With a typical depth of AB $\sim 21.25$\,mag per band, this filter set thus allows for an indiscriminate and accurate characterization of the stellar population in our Galaxy, it provides an unprecedented 2D photo-spectral information for all resolved galaxies in the local universe, as well as accurate photo-$z$ estimates (at the $\Delta\,z\sim 0.01$--$0.03$ precision level) for moderately bright (up to $r\sim 20$\,mag) extragalactic sources. While some narrow band filters are designed for the study of particular emission features ([OII]/$\lambda$3727, H$\alpha$/$\lambda$6563) up to $z < 0.015$, they also provide well-defined windows for the analysis of other emission lines at higher redshifts. As a result, J-PLUS has the potential to contribute to a wide range of fields in Astrophysics, both in the nearby universe (Milky Way structure, globular clusters, 2D IFU-like studies, stellar populations of nearby and moderate redshift galaxies, clusters of galaxies) and at high redshifts (emission line galaxies at $z\approx 0.77, 2.2$ and $4.4$, quasi stellar objects, etc). With this paper, we release $\sim$36\,deg$^2$ of \jplus data, containing about $1.5\times 10^5$ stars and $10^5$ galaxies at $r<21$\,mag. These numbers are expected to rise to about 35 million of stars and 24 million of galaxies by the end of the survey.}
\titlerunning{\jplus}
%
\authorrunning{A.~J.~Cenarro et al.}
\maketitle
%
%
%
%

\section{Introduction}
\label{sec:Intro}

The success of large sky, broad band, photometric surveys like, e.~g., the Sloan Digital Sky Survey \citep[SDSS,][]{Sloan_survey_York00} demonstrated that multi-filter surveys can provide crucial information on the spectral energy distributions (SED) for millions of astronomical objects. This is particularly evident for some science cases in which spectral resolution is not a limiting factor and, hence, multi-object spectroscopy is not critically required. One of the major advantages of photometric surveys relies on the fact that the object selection is not biased by any prior other than the limiting magnitude in each band, hence favoring completeness and making possible the generation of more massive data sets. 

In recent years, a growing number of multi-filter surveys with larger numbers of photometric bands like, e.g., COMBO-17 \citep{Wolf2003}, ALHAMBRA \citep{Moles2008}, COSMOS \citep{Ilbert2009}, MUSYC \citep{Cardamone2010}, CLASH \citep{postman12}, SHARDS \citep{PerezGonzalez2013}, have opened new ways to shed light on many areas in Astrophysics and Cosmology. These surveys share in common the use of a set of broad, intermediate and/or narrow band filters providing sufficient spectral information that enables addressing certain scientific cases without the need to resort to spectroscopy. The success of such surveys has motivated the design of new, challenging projects like, e.g., the Physics of the Accelerating Survey \citep[PAU,][]{pau_survey}, the Javalambre-PAU Astrophysical Survey \citep[\jpas,][]{Benitez2014}, and the Javalambre-Photometric Local Universe Survey, hereafter \jplus\footnote{\url{http://www.j-plus.es}}, whose presentation is the main goal of this paper.

As we explain next, the motivation for a survey like \jplus lies in the context of the \jpas survey. \jpas is a very wide field astrophysical survey to be developed at the Observatorio Astrof\'{\i}sico de Javalambre \citep[OAJ;][]{2010SPIE.7738E..0VC,2012SPIE.8448E..1AC,2014SPIE.9149E..1IC}. \jpas will be conducted with the Javalambre Survey Telescope, JST/T250, a large-etendue 2.5\,m telescope, and JPCam, a panoramic $4.7\deg^2$camera with a mosaic of 14 large format CCDs, amounting to 1.2\,Gpix. The main scientific driver of \jpas is the measurement of radial Baryonic Acoustic Oscillations with an unprecedented accuracy. To achieve this, \jpas is planned to observe around $8\,500\deg^2$ of the Northern Sky with a set of 54 contiguous, narrow band optical filters (145\,\AA\ width each, placed $\sim 100$\,\AA\ apart), plus two broad band filters at the blue and red sides of the optical range, and 3 SDSS-like filters. The filter set is designed to provide 0.3\,\% relative error ($0.003\times[1+z]$) photometric redshifts (photo-$z$s) for tens of millions of luminous red and emission line galaxies (ELGs), plus about two million quasi-stellar objects (QSOs) with similar photo-$z$ quality, sampling an effective volume of $\sim$14\,Gpc$^3$ up to $z=1.3$. In addition, \jpas is expected to detect hundreds of thousands of galaxy clusters and groups; produce one of the most powerful cosmological lensing surveys; and detect, classify and measure redshifts for thousands of SNe at the $1$\,\% precision level \citep{xavier2014}. All these observations should provide complementary constraints on Dark Energy, each of them suffering from different systematics, and thus providing a solid description of the nature of Dark Energy. 
Beyond these cosmological goals, the key to the \jpas potential is its innovative approach. The use of the narrow band filters makes \jpas to be equivalent to a low resolution Integral Field Unit (IFU) of the Northern Sky, providing the SED of every pixel of the sky and, ultimately, a 3D image of the Northern Sky with an obvious wealth of potential astrophysical applications\footnote{One must bear in mind that, unlike in standard IFU spectroscopy, in the approach adopted by \jplus and \jpas, data corresponding to different bands/spectral ranges may be acquired at different times. For this reason, the term ``hyperspectral imaging" has also been proposed to dub the non-simultaneous, multi narrow, medium-width, and broad band photometry conducted by \jplus. }.

The \jplus project is envisioned as a multi-filter survey aiming to support and complement \jpas at both the technical and scientific levels. With a unique set of 12 broad, intermediate and narrow photometric bands in the optical range, \jplus was conceived to be implemented at the OAJ, using an 83\,cm telescope with a large field of view (FoV) and a very efficient, low read-out noise panoramic camera. Telescope, camera and filters are specifically designed under the main goal of conducting \jplus, hence optimizing the whole system to achieve the maximum outcome and efficiency of the survey. 

\jplus' original goal to support the photometric calibration for \jpas drove the definition of the \jplus filters. These filters have been conceived to retrieve accurate stellar SEDs and assign stellar templates to all the stars brighter than a certain magnitude. Those templates will ultimately allow to compute, for each star, synthetic magnitudes for the 59 \jpas photometric bands in advance to \jpas observations. With several dozens of stars in each of the 14 large format CCDs of JPCam it will be possible to compute the photometric zero points for each exposure of \jpas \citep[see, e.g.,][]{varela2014}. This calibration technique, together with alternative methods based on stellar locus in multiple \jpas color-color diagrams, guarantees the success of the \jpas calibration to uncertainties below 2\,\% in each filter.

In addition, the timing of \jplus is also an important asset. From the beginning, \jplus was thought to be conducted well in advance of \jpas. First, because \jplus aims to observe most of the \jpas footprint beforehand to support and complement the photometric calibration of \jpas and its science. Second, because most of the technical developments related with the \jplus deployment (e.g., data reduction, pipelines, calibration procedures, scientific software, storage hardware, etc.) 
constitute the natural basis for \jpas, in many cases requiring only minor modifications to be adapted to the \jpas specifications. 

The details of the photometric calibration of \jpas using \jplus data will be addressed in a dedicated paper (Varela et al. in preparation). In this paper, however, we focus on the scientific value of \jplus. There are two main primary goals related with local Universe science that have been prioritized during the survey definition. The first one is the production of multi-filter photometric information for millions of Milky Way (hereafter MW) halo stars in the \jpas footprint. This is a very valuable data set for multiple scientific cases in stellar astrophysics, like, e.g., the determination of atmospheric parameters, the detection of extremely metal poor stars, the study of brown dwarfs, variable sources, etc.  The second primary goal is to optimally exploit the IFU-like science that \jplus can provide at very low spectral resolution. In particular, \jplus is aimed to focus on the detailed characterization of 2D stellar populations, star formation rates, and the role of environment for tens of thousands of galaxies in the local Universe ($z < 0.05$). In the context of these two main goals, the \jplus filter set has been defined to be particularly sensitive to rest-frame spectral features that are key to i) characterize the different stellar spectral types, luminosity classes and stellar metallicities, ii) disentangle ages and metallicities from the integrated light of local galaxies, and iii) determine reliable strengths of nebular emission lines such as H$\alpha$ and [OII]/$\lambda$3727. 

The characteristics and strategy of the survey have been modulated to extract the maximum scientific return in many other areas of Astrophysics. \jplus will observe a significant fraction of the Northern sky in 12 photometric optical bands, hence providing multi-filter information for every pixel in the survey's footprint. This allows to address a very wide range of scientific cases, essentially all those that are benefited by i) the very large sky coverage of the survey; ii) the use of specific filters, some of them with narrow band passes located at key rest-frames spectral features; iii) a massive volume of data for most categories of astronomical objects, and iv) the time domain, as a result of the specific strategy to observe each field in the different filters at different times plus the revisit of fields observed under poor atmospheric conditions.

Apart from describing the \jplus mission, the main goal of the present paper is to illustrate some of the most relevant scientific capabilities of \jplus based on real data. With this aim, in this paper we make public the first data release of the project, which amounts to $\sim 50$\,deg$^2$ of \jplus data split in two categories: $\sim 36$\,deg$^2$ in the \jplus footprint, constituting the so called \jplus Early Data Release, and $\sim 14$\,deg$^2$ in particularly interesting regions of the sky that were observed as part of the \jplus Science Verification Data. This data set is representative of the whole \jplus survey and allows to provide the reader with a flavor of the most relevant science cases that \jplus will address: MW science, stellar clusters, stellar populations of nearby and moderate redshift galaxies ($z < 0.015$), emission line studies, 2D IFU-like studies, high redshift galaxies, QSOs, clusters of galaxies, etc. 

\jplus has already collected data over more than 1\,000\,deg$^2$ of data, which should become publicly accessible during 2018. In the meantime, the {\it South-Photometric Local Universe Survey} (S-PLUS) is scanning the Southern sky from Cerro Tololo with identical telescope, camera and filter set as \jplus, thus strengthening the large scale character of both projects. 

Overall, the paper is structured as follows. Section~\ref{sec:description} introduces a general description of the survey making emphasis on the technical framework of \jplus: the OAJ, the JAST/T80 telescope and its panoramic camera T80Cam, the \jplus filter set, the survey strategy, the data reduction pipeline and, finally, the photometric calibration. Section~\ref{sec:J-PLUS_EDR} is devoted to describe the \jplus' Early Data Release and Science Verification data sets that are publicly released with this paper, on the basis of which Sect.~\ref{sec:J-PLUS_science} illustrates briefly some of the main science cases that will be covered by \jplus. Finally, the summary and conclusions are presented in Sect.~\ref{sec:conclusions}.

\section{\jplus definition and implementation}
\label{sec:description}

\jplus is a dedicated multi-filter survey of $\sim 8\,500$\,deg$^2$ that will be conducted from the OAJ during the next years using the JAST/T80 telescope. JAST/T80 is an 83\,cm telescope with a large FoV, mounting T80Cam, a panoramic camera that provides a FoV of $2\,\deg^2$, and a set of 12 broad, medium and narrow-band filters. This filter set has been particularly defined to be sensitive to key stellar spectral features in the rest frame, thus being optimal for MW science and stellar population studies of galaxies in the local Universe. In addition, the survey strategy has been fine-tuned to optimize the scientific return in a wide range of applications in many other areas of Astrophysics. 

In this section we describe the main technical characteristics of the survey, starting with the observatory (Sect.~\ref{sec:oaj}), telescope and camera (Sect.~\ref{sec:T80}), and filters (Sect.~\ref{sec:filters}), to continue with the survey strategy (Sect.~\ref{sec:J-PLUS_strategy}), data reduction (Sect.~\ref{sec:J-PLUS_pipelines}), and calibration (Sect.~\ref{sec:J-PLUS_calibration}).

\subsection{The OAJ: an observatory for large sky surveys}
\label{sec:oaj}

The OAJ\footnote{\url{http://www.oaj.cefca.es}} is an astronomical facility
located at the Pico del Buitre of the Sierra de Javalambre, in Teruel,
Spain. The site, at an altitude of $1957$\,m, has excellent astronomical
characteristics in terms of median seeing (0.71\,arcsec in V band, with
a mode of 0.58\,arcsec), fraction of clear nights ($53\,\%$ totally clear,
$74\%$ with at least a $30\%$ of the night clear) and darkness, with a typical sky surface brightness of $V\sim 22$\,mag\,arcsec$^{-1}$ at zenit during dark nights, a feature quite exceptional in continental Europe. Full details about the site testing of the OAJ can
be found in \cite{2010PASP..122..363M}.

The OAJ was defined, designed and constructed to carry out large sky
surveys with dedicated telescopes of unusually large FoVs. The two main telescopes at the OAJ are the Javalambre Survey Telescope (JST/T250), a 2.55\,m telescope with 3\,deg diameter FoV, and the Javalambre Auxiliary Survey Telescope (JAST/T80), an 83\,cm telescope with a FoV diameter of 2\,deg. JAST/T80 is the telescope dedicated to the development of \jplus, whereas J-PAS will be carried out at the JST/T250. 

The definition, design, construction, exploitation and management of the observatory and the data produced at the OAJ are responsibility of the Centro de Estudios de F\'{\i}sica del Cosmos de Arag\'on (CEFCA\footnote{\url{http://www.cefca.es}}). The OAJ project started in March 2010, mostly funded by the {\it Fondo de Inversiones de Teruel}, a programme supported by the local Government of Arag\'on and the Government of Spain, and is essentially completed since 2015. In October 2014, the OAJ was awarded with the recognition of Spanish ICTS ({\em Infraestructura
Cient\'{\i}fico T\'ecnica Singular}) by the Spanish Ministry of Economy and Competitiveness.

\subsection{JAST/T80 and T80Cam}
\label{sec:T80}

\begin{figure*}[t!] \centering
\includegraphics[width=7in]{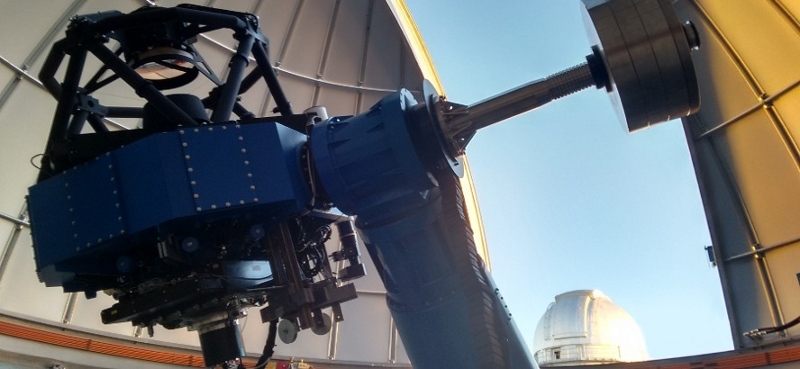}
\caption{View of the JAST/T80 telescope inside its dome at the Observatorio Astrof\'{\i}sico de Javalambre.} \label{fig:JAST_T80} \end{figure*}

The JAST/T80 is an 83\,cm telescope with a FoV of 2\,deg diameter and 
fast optics (F$\#$4.5), which drives a plate scale at the Cassegrain
focal plane of 55.56\,arcsec\,mm$^{-1}$. JAST/T80 has a
German-equatorial mount. Figure~\ref{fig:JAST_T80} illustrates JAST/T80
inside the 6.2\,m dome building at the OAJ. The optical tube assembly
has also a very compact layout, with just $826$\,mm between M1 and M2.
With an overall weight of around 2500\,kg, JAST/T80 supports instruments
at its Cassegrain focus of up to 80\,kg.

\begin{table} \caption{Main technical characteristics of the JAST/T80
telescope} \label{tab:JAST/T80_technical} \centering \begin{tabular}{r l
} 
\hline\hline Mount: & German equatorial\\ Optical configuration: &
Ritchey Chr\'etien modified\\ M1 diameter: & 83\,cm \\ Field corrector:
& 3 spherical lenses \\ Effective collecting area: & 0.44\,m$^2$\\
Focus: & Cassegrain\\ F$\#$: & 4.5\\ Focal length:& 3712\,mm\\ Plate
scale: & 55.56\,arcsec\,mm$^{-1}$\\ FoV (diameter) & 2.0\,deg \\
Etendue: & 1.5\,m$^2$deg$^2$\\ EE50 (diameter)    & $9\,\mu{\rm m}$\\
EE80 (diameter)    & $18\,\mu{\rm m}$ \\ \hline \end{tabular}
\end{table}

The optical design of the JAST/T80 is based on a modified
Ritchey-Chr\'etien configuration, including a field corrector of three
spherical lenses of fused silica. The lens diameters are in the range
$152$--$170$\,mm. The whole system is optimized to work in the optical
range, from 330 to 1100\,nm, yielding a polychromatic image quality
better than $9.0\,\mu$m (EE50; diameter) inside the 13\,cm diameter
focal plane ($3.1$\,deg$^2$), after having considered all error
sources in the error budget. The design is also optimized to account for
the \jplus filters and the T80Cam entrance window in the optical path.
Table~\ref{tab:JAST/T80_technical} illustrates a summary of the main
technical characteristics of JAST/T80.

Because of the large FoV and fast optics, the JAST/T80 M2 is
controlled actively with an hexapod that allows to perform small M2
corrections, keeping the system in focus and free of aberrations all
over the FoV. This is done through a wave-front curvature sensing
technique developed at CEFCA \citep[see][]{2012SPIE.8450E..0IC} that
computes the optimal hexapod position for a given temperature and
telescope pointing. This technique makes use of the scientific CCD of
T80Cam, demanding only a few minutes every few hours of
observation. In between such computations, M2 corrections are applied
automatically in closed loop during the survey execution according to
an empirically calibrated M2 control law. This control law is built
on the basis of hundreds of telescope positions and temperatures.  To
improve the efficiency of the system, during the day time the JAST/T80
dome is air-conditioned at the expected temperature of the coming
night to minimize temperature gradients in the telescope and camera.

Since the overall effective etendues of the OAJ systems are ultimately
determined by the CCD filling factor at the telescope focal plane, the
OAJ instrumentation is designed to take full advantage of the large FoV
of the telescopes and the seeing conditions of the site. In this sense,
JAST/T80 is equipped with an only instrument, T80Cam, a panoramic camera
that will remain mounted at the telescope during the entire
execution of \jplus, hence minimizing overheads in the operation due to
instrument exchanges.

T80Cam includes a grade-1, back-side illuminated, low-noise $9.2\,{\rm
kpix}\times9.2$\,kpix CCD of 10$\,\mu$m\,pix$^{-1}$, specifically developed
by Teledyne e2v (UK) for the \jplus and J-PAS projects. T80Cam provides a
usable FoV of 2\,deg$^2$ with a plate scale of 0.55\,arcsec\,pix$^{-1}$.
This full wafer CCD is read simultaneously from 16 ports, achieving
read-out times of $12$\,s with a typical read-out noise of $3.4\,{\rm
e}^{-}$ (RMS). It has an image area of $92.16\,{\rm mm} \times
92.32\,{\rm mm}$ and a broadband anti-reflective coating for optimized performance in
the range $380$--$850$\,nm.

T80Cam is a direct imaging camera designed to work in fast convergent
beam at the Cassegrain focus of the JAST/T80, not having additional
optical elements other than the \jplus filters and the cryostat
entrance window, which is optically powered. In fact, the camera
entrance window is the fourth element of the JAST/T80 field corrector, that together with the filters has been accounted for when defining the optical design
of JAST/T80.  The window is a 10\,mm thick, weakly powered
field-flattener with an 8\,mm distance between its inner surface and the
focal plane.

\begin{figure*} \centering
\includegraphics[height=1.4in]{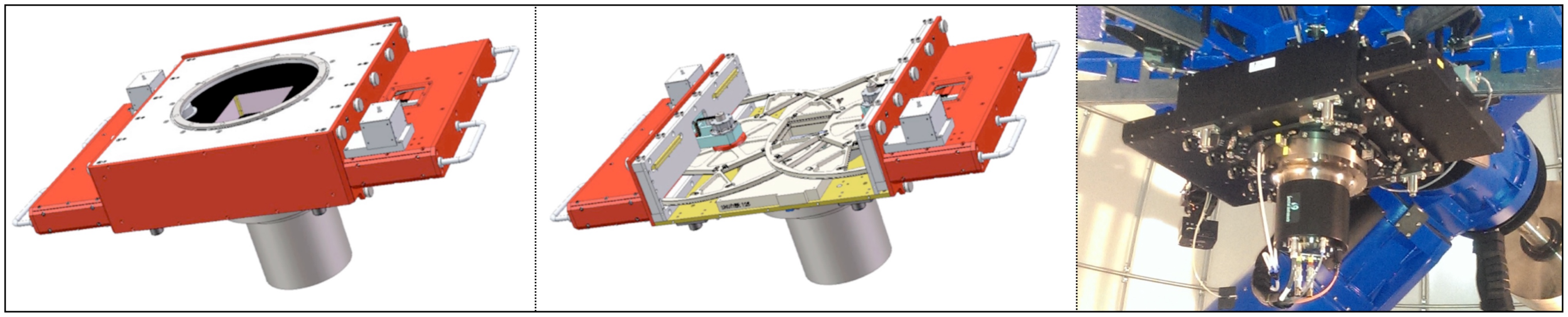}
\caption{{\sl Left:} T80Cam 3D model design completely assembled. The
top part of the instrument represents the filter and shutter unit,
containing the shutter and the two filter wheels. The gray,
cylindrical-shaped object underneath the filter and shutter unit
represents the cryogenic camera. {\sl Center:} the same view of T80Cam
model after cover removal, showing the two filter wheels and the
shutter. {\sl Right:} T80Cam integrated at the Cassegrain focus of the
JAST/T80 telescope.} \label{fig:T80Cam} \end{figure*}

T80Cam (Fig.~\ref{fig:T80Cam}; left) consists of two main subsystems, namely
the filter and shutter unit (FSU) and the cryogenic camera. Apart from
the mechanical flange to attach the instrument to the telescope, the FSU
includes a two-curtain shutter provided by Bonn-Shutter UG (Germany)
that allows taking integration times as short as 0.1\,s with an
illumination uniformity better that $1\%$ over the whole FoV and the
filter unit. The latter consists of two filter wheels with 7 positions
each (Fig.~\ref{fig:T80Cam}; center), allowing to host all the 12 \jplus
filters (6 filters plus an empty space in each filter wheel). This avoids 
the need to exchange filters during observations or from night to night,
thus minimizing risks and maintenance downtimes. The camera system is a
1110S camera manufactured by Spectral Instruments (USA).
Figure~\ref{fig:T80Cam} (right) shows a view of T80Cam mounted on the Cassegrain focus of the JAST/T80. The cryogenic camera system consists of the cryostat, the
powered entrance window, the science CCD, its control electronics, and
the vacuum and cooling systems. The sensor is cryo-cooled to an
operating temperature of $-100^\circ\,{\rm C}$ with a cryo-tiger
refrigeration system, a closed-cycle Joule-Thomson effect cryogenic
refrigerator device. The chamber is evacuated to
$10^{-4}$\,Torr using a turbo dry vacuum pump.

A summary of the technical characteristics of T80Cam is presented in
Table~\ref{tab:T80Cam_technical}. In addition, full technical and
managerial details on T80Cam can be found in \cite{2012SPIE.8446E..6HM}
and \cite{2015IAUGA..2257381M}.

\begin{table} \caption{Main technical performances of the T80Cam
panoramic camera at the JAST/T80 telescope} \label{tab:T80Cam_technical}
\centering
\begin{tabular}{r l } \hline\hline CCD format	     &
$9216\times9232$\,pix \\ & $10\,\mu{\rm m}\,{\rm pix}^{-1}$ \\ Pixel
scale	     & $0.55$\,arcsec\,pix$^{-1}$ \\ FoV coverage &
$2.0\deg^2$ \\ Read out time	    & 12\,s \\ Read out noise	     &
$3.4\,{\rm e}^{-}\,{\rm pix}^{-1}$ \\ Full well & $123\,000\,{\rm
e}^{-}$ \\ CTE & 0.99995 \\ Dark current	    & $0.0008\,{\rm
e}^{-}\,{\rm pix}^{-1}\,{\rm s}^{-1}$ \\ Number of filters  & 12 \\
\hline \end{tabular} 
\end{table}

\subsection{\jplus filter set}
\label{sec:filters}

The main goals of \jplus hinge on the accurate determination of SEDs of MW stars and nearby galaxies. It is therefore clear that the \jplus filter set must be sensitive to both the optical continuum and the most prominent spectral features. In this sense, it has been widely demonstrated that it is possible to retrieve reliable stellar SEDs with a set of around 10 to 15 intermediate-broad band optical filters centered at key spectral regions \citep[e.g.,][]{2004A&A...419..385B,2006MNRAS.367..290J,carrasco2006}.

Following this strategy, the \jplus filter set consists of the 12
filters defined in Table~\ref{tab:JPLUS_filters}. Among them, 4 are SDSS filters \citep[$g,
r, i$, and $z$;][]{1996AJ....111.1748F}, providing the low
frequency continuum, while 6 are intermediate band filters of $200$--$400$\,\AA\
width, centred on key absorption features. These are the $u$ filter, in common with J-PAS and located at the blue side of the $3\,700$--$4\,000\,$\AA\ Balmer break region, and the filters $J0395$ for Ca H$+$K, $J0410$ for H$\delta$, $J0430$ for the G-band, $J0515$ for the Mg$b$
triplet, and $J0861$ for the Ca triplet. The \jplus filter set is completed
with 2 narrow band filters, $J0378$ and $J0660$, also in common with the J-PAS filter set. 
These two filters are sensitive to the [OII]/$\lambda3727$ and
H$\alpha$/$\lambda6563$ emission lines, respectively. The three filters in common
with J-PAS are envisioned as an added value for the overall calibration
procedure, as they will allow second order corrections of the zeropoints
to anchor the J-PAS calibration.

\begin{table}[ht] 
\caption{The \jplus filter system. Comments: (a) In common with J-PAS; (b) SDSS. Rest-frame key spectral features matching the location of narrow and mid band filters are also indicated.} 
\label{tab:JPLUS_filters}
\centering 
	\begin{tabular}{c c c l } 
	\hline\hline 
                 & Central        &   		&           \\ 
       Filter  & Wavelength       & FWHM & Comments\\ 
	       	& [\AA]             & [\AA] 			&         \\ 
	\hline
	$u$	 	&3485 	&508		& (a) \\ 
	$J0378$ 	&3785 	&168 	& [OII]; (a)\\ 
	$J0395$ 	&3950	&100		& Ca H$+$K\\ 
	$J0410$ 	&4100	&200		& H$_\delta$\\ 
	$J0430$	&4300 	&200		& G-band\\ 
	$g$		&4803 	&1409	& (b)\\ 
	$J0515$ 	&5150	&200		& Mg$b$ Triplet\\ 
	$r$ 		&6254	&1388	& (b)\\ 
	$J0660$ 	&6600 	&138		& H$\alpha$; (a)\\ 
	$i$		&7668 	&1535	& (b)\\ 
	$J0861$	&8610 	&400		& Ca Triplet\\ 
	$z$ 		&9114	&1409	& (b)\\ 
	\hline 
\end{tabular}
\end{table}

\begin{figure*} \centering
\includegraphics[width=7in]{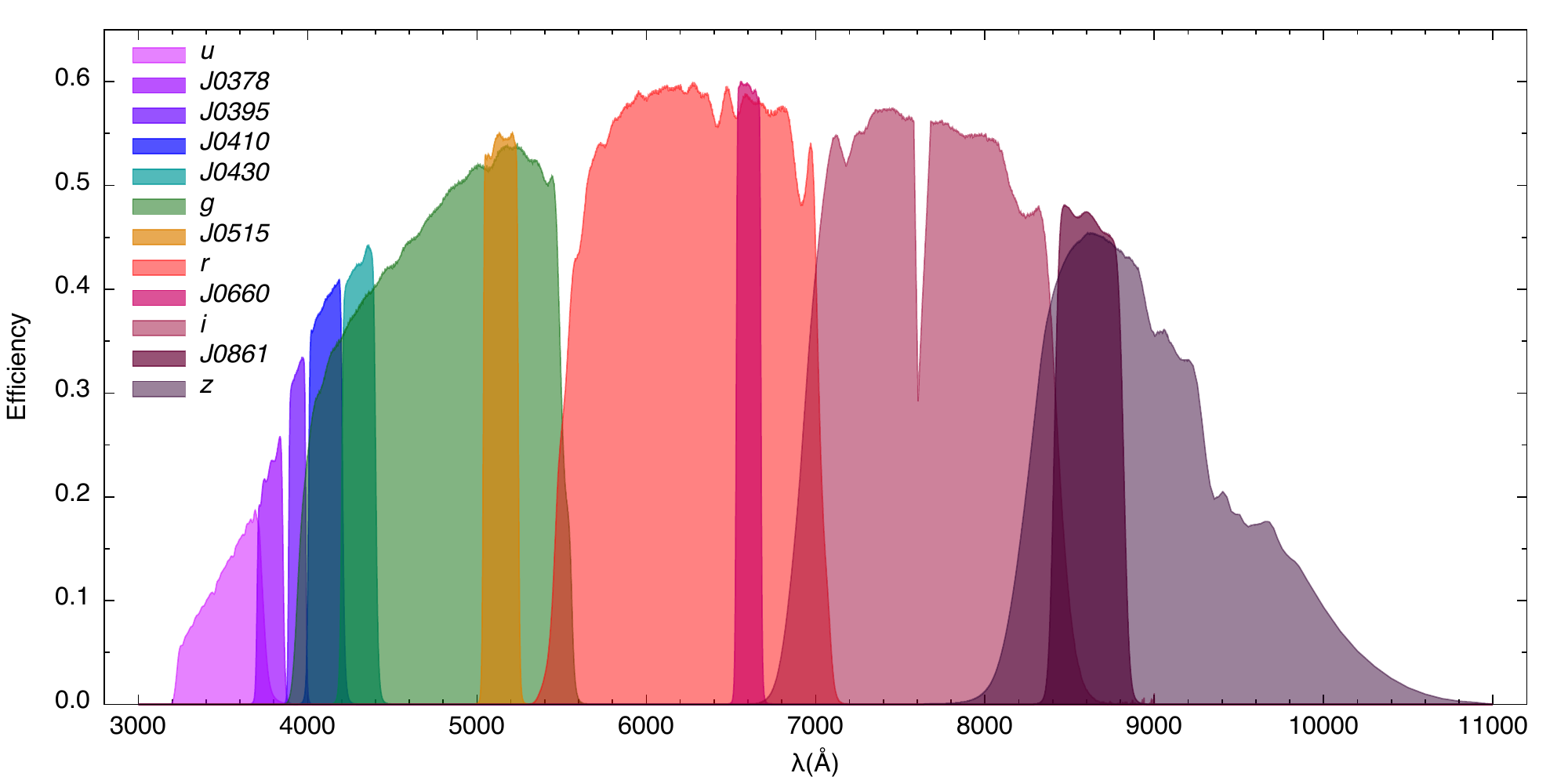} 
\caption{Efficiency curves measured for the set of 12 \jplus filters, including the effect of the entire system (sky, mirrors, lenses, and CCD).
}
\label{fig:JPLUS_filters} \end{figure*}

The \jplus filters have been manufactured by SCHOTT (Switzerland). A
detailed technical description of the filter requirements and design can be found in
\cite{2012SPIE.8450E..3SM}. Fig.~\ref{fig:JPLUS_filters} shows the
final efficiency curves of the filters.

The filter arrangement is tuned to provide scientifically valuable
data for many fields of Astrophysics. Since the intermediate band
filters are sensitive to the strengths of key features of old stellar
populations, \jplus is very well suited for analyzing the stellar
populations of nearby galaxies up to a limiting redshift of
  $z\sim 0.015$ (set by the typical filter width). In
addition, the two narrow band filters are ideal for mapping the star
formation rates in nearby galaxies in the range $0 < z < 0.05$. These
considerations are fully addressed in Sect.~\ref{sec:J-PLUS_science}.

\subsection{\jplus survey strategy}
\label{sec:J-PLUS_strategy}

As explained in Sect.~\ref{sec:Intro}, one of the main goals of \jplus
is to provide reliable stellar SEDs for millions of stars in the MW
halo. For this goal, simulations reveal that it is sufficient reaching
signal-to-ratio SNR$\geq 50$ for all stars brighter than $\sim 18$\,mag\footnote{Unless explicited otherwise, magnitudes in this work will refer to the AB system.} in each \jplus
filter. Given the T80Cam FoV of 2\,deg$^2$, there are more
than one thousand stars per pointing and filter that reach $\sim 18$\,mag
in the Galactic halo. In terms of limiting magnitude, guaranteeing the
above numbers is approximately equivalent to reaching $\sim 21.0$\,mag for
SNR$\geq 3$ for point-like sources, in a circular aperture of
3\,arcsec.  This depth suffices to guarantee unprecedented
IFU-like science for thousands of nearby galaxies, both for 2D stellar
populations gradients and SFR studies.

\begin{table}[ht]
\caption{Summary of the \jplus limiting magnitudes (for SNR$>3$) and zero point calibrations. Two different quotes are given for the uncertainties in the zero point calibrations, one from EDR and another one from the amount of \jplus data collected until November 2017, after using the stellar locus and minimization in overlap areas in $u,g,r$, and $i$ filters.}
\label{tab:J-PLUSmodes}
\centering
\begin{tabular}{cccccc}
\hline
\hline
\\
Filter  	&  $m_{\rm lim}^{\rm J-PLUS}$ & $m_{\rm lim}^{\rm EDR}$  & $\langle {\rm ZP} \rangle$  & $\sigma_{\rm ZP}^{\rm EDR}$ & $\sigma_{\rm ZP}^{\rm J-PLUS}$\\
\hline
$u$    		&  21.00 	&  21.6		&  21.13		& 0.037 & 0.023\\	
$J0378$		&  21.00 	&  21.5		&  20.54		& 0.084 & 0.026\\
$J0395$		&  21.00 	&  21.4		&  20.32		& 0.072 & 0.026\\	
$J0410$		&  21.25 	&  21.5		&  21.30		& 0.058 & 0.018\\	
$J0430$		&  21.25 	&  21.5		&  21.37		& 0.050 & 0.018\\	
$g$		&  22.00 	&  22.2		&  23.58		& 0.042 & 0.013\\	
$J0515$		&  21.25 	&  21.4		&  21.52		& 0.039 & 0.012\\
$r$             &  22.00 	&  21.9		&  23.52		& 0.039 & 0.010\\
$J0660$     	&  21.25 	&  21.3		&  21.04		& 0.042 & 0.012\\
$i$	        &  21.75 	&  20.8		&  23.25		& 0.042 & 0.012\\	
$J0861$     	&  20.50 	&  20.8		&  21.54		& 0.052 & 0.017\\	
$z$		&  20.75 	&  20.5		&  22.63		& 0.024 & 0.017\\
\hline
\end{tabular}
\end{table}

The exposure times of \jplus are set to reach the
required signal-to-noise at {$\sim 18$\,mag}. Since \jplus fields are observed at different moon phases and moon distances, the optimal exposure time will
depend on the brightness of the sky. For this reason, during
commissioning we have empirically modelled the dependence of the sky
brightness with respect to both distance and phase of the Moon.
The estimated sky brightness at every \jplus pointing is then
  given as input to the exposure time calculator, which determines the
  particular exposure time for that pointing and time. Given the
change in background, observing in the 12\,filters takes typically
about 35\,minutes in dark time, and about 1.5\,hours in bright time.

With the goal of observing the same $\sim 8\,500$\,deg$^2$ in the footprint of \jpas, \jplus is scheduled in $\sim 4\,250$ pointings. For the most immediate observations, three priority areas have been selected amounting to about 1\,500\,deg$^2$:
\begin{itemize}
	\item Priority Area -- North 1 (PAN1): Defined by 120\,deg < RA < 180\,deg and 30\,deg < DEC < 42\,deg, this area (580\,deg$^2$) overlaps with eROSITA-Germany\footnote{\url{http://www.mpe.mpg.de/eROSITA}} and UKIDSS-LAS-1\footnote{\url{http://www.ukidss.org}} fields.
	\item Priority Area -- North 2 (PAN2): Defined by 180\,deg < RA < 245\,deg and 42.5\,deg < DEC < 57\,deg, this area (610\,deg$^2$) overlaps with HETDEX\footnote{\url{http://hetdex.org}}, ELAIS-N1, and ALHAMBRA-6\footnote{\url{http://www.alhambrasurvey.com}} fields.
	\item Priority Area -- South (PAS): Defined by 0\,deg < RA < 42.2\,deg and -5\,deg < DEC < 8\,deg, this area (550\,deg$^2$) overlaps with the SDSS Stripe 82\footnote{\url{http://cas.sdss.org/stripe82/en/}}.
\end{itemize}
In general, each field is observed when its observability is optimal, typically within the requirements of airmass lower than 1.5 and seeing better than 1.5\,arcsec. The ultimate observing sequence is managed by the \jplus Scheduler and Sequencer according to predefined observability, efficiency and image quality criteria.

\subsection{\jplus pipelines and data management}
\label{sec:J-PLUS_pipelines}

The \jplus data are handled and processed using the data management
software developed by the Data Processing and Archiving Unit ({\it
  Unidad de Procesado y Archivo de Datos}; hereafter UPAD) of CEFCA.
The technical details of \jplus data handling and data processing will
be explained in a forthcoming article. Here we simply describe some of
the steps involved in the treatment of the \jplus data presented in
this paper.

During the execution of \mbox{J-PAS} and \mbox{\jplus}, the OAJ
telescopes will produce thousands of images per night, amounting to
several terabytes of data during a single, typical, observing night. To
  digest such a data rate, the pipelines have been designed to work
automatically, i.e., with no human supervision. The OAJ hosts a
  set of dedicated servers that manage the transmission of the images
from the camera servers to the OAJ storage and processing unit, where
an online processing is done. This first processing uses a
  lightweight version of the pipeline, for real time visual inspection
  and first order diagnosis. Also, right after image acquisition, raw
data are automatically transferred to the UPAD at CEFCA for archival
and a complete processing. This is done via a dedicated radio-link
that connects the OAJ with CEFCA headquarters.
 
  Once the raw data reach the main UPAD archive, they are
  automatically uploaded to a management database which stores all the
  metadata of the collected images. This database also stores all the
  processes done on the data, and is used to control whether the
  inputs for each stage are ready, and to automatically prepare the
  jobs to be executed. On a daily basis, the Data Management Software
  automatically processes the data collected during the last night in
  order to check their quality. It next updates the survey databases
  and feeds the Scheduler to compute the telescope targets of the
  following nights. Eventually, in order to optimize the scientific
  quality of the data, the pipelines reprocess all the images with the
  optimal master Calibration Frames (CFs) once they are available. 
 
 The image processing pipelines have three main stages. 
 The first one is related to the generation of master CFs, which are used 
 to correct for the instrumental imprint on the individual images at a later stage.  
 These CFs consist of a master bias, the master dome or sky flats, a fringing
  pattern, and the illumination correction maps. To compute any of
  those CFs the pipeline requires as input the time interval and the
  instrumental configuration (telescope, camera and detector
  parameters, read-out configuration, etc). With this information in
hand, the pipeline queries the database for the needed images and
processes the master CFs. Once the CFs are generated, they are made
available, through a web service, together with relevant information
that allows to assess their quality. Finally, the master CFs require
human validation to be used in further processing of science images.

The second stage is related to the processing of the different
individual images. The pipeline corrects the instrumental signature
using the validated CFs. The corrections (bias, flat fielding,
fringing, etc.) that need to be applied to each image type are
  stored in the configuration files. Since the proper CFs for
  each image depend upon the required corrections, the observing epoch
  and the instrumental configuration, the pipeline must query the
  management database in order to assign the correct CFs. The
pipeline also computes and assigns cold and hot pixel masks to every
image, while generating masks for other contaminants (e.g., cosmic rays and
satellite traces). If required, the pipeline can also interpolate the
masked areas. For the detection and masking of cosmic ray hits, the
pipeline uses an implementation of {\tt L.A.Cosmic}
\citep{2001PASP..113.1420V}. During the processing of the individual
images, the pipeline calibrates the astrometry and the photometry
\citep{2006ASPC..351..112B}, and generates a PSF map and a master PSF
for each individual frame. Once the individual images are processed, a
first catalogue is created over the individual images using {\tt
  SExtractor} \citep{1996A&AS..117..393B}.

At a final stage, the pipeline combines the processed images from the same sky area to provide a deeper, photometric calibrated image for each filter. The pipeline makes use of the packages {\tt Scamp} \citep{2006ASPC..351..112B} and {\tt Swarp} \citep{2002ASPC..281..228B} to perform the astrometric calibration and image coadding. The combined images for a given tile in the whole set of filters constitute a datacube. {\tt SExtractor} source catalogues are extracted from the final tiles and datacubes and are stored in an internal database system. Two kind of catalogues are produced. In the first one, the source detection and segmentation are done independently for the combined images in each filter. In the second one, the detection and aperture definition is done in the $r$ band, which is used as a reference image. Prior to computing the final catalogues, the PSF of each filter image is homogenized to match the one of the $r$ filter, defined as reference filter in \jplus.
This technique was already used to produce the photometric catalogues for the ALHAMBRA survey \citep{2014MNRAS.441.2891M}.

The archival and processing of the J-PAS and \jplus data is done in a dedicated data center. The data center infrastructure and processing software are designed to handle the enormous data flow produced by the OAJ panoramic cameras, and specially to minimize the time required to transfer the images to the processing nodes, and to read/write data from/to disks. The main storage uses two different technologies. In order to feed data to the processing nodes, the data that are accessed frequently by the pipelines are kept in a disk storage cluster. The storage cluster runs a distributed file system providing access to the data through several servers, and together with the core network they provide an aggregated bandwidth above $50$\,Gbps to the processing servers. The permanent archive and the backup of the data is done in a robotic tape library.  To process the \jplus images, the UPAD hosts dedicated nodes that have a large RAM to operate with the images without writing intermediate products to disk. The UPAD also counts with a large internal disk scratch used to store copies of the frequently required data, such as the CFs. More information on the facilities employed to process the \jplus data is provided in \citet{cristobal2014}.

\subsection{\jplus photometric calibration}
\label{sec:J-PLUS_calibration}

The photometric calibration of \jplus faces two main challenges, namely the variety of observational conditions in which, throughout the project, \jplus images will be taken, and the use of a unique set of purposely defined filters. However, the difficulty of these tasks is to some extent alleviated by the large amount of external data made available by projects like SDSS \citep{Sloan_survey_York00}, PanSTARRS \citep{2002SPIE.4836..154K} and Gaia \citep{gaia}.

In this context, with the ultimate goal of being able to calibrate \jplus at the widest variety of observing conditions, we choose to apply a battery of calibration procedures rather than relying on a single calibration technique. While the details of the calibration procedures will be presented in an upcoming work (Varela et al, in preparation), here we briefly outline the calibration procedures being currently applied on current \jplus data:

\begin{itemize}
\item\textit{SDSS spectroscopy}. This is simply done by convolving the SDSS spectra with the spectral response for each filter of \jplus, yielding synthetic magnitudes whose comparison to the observed one provides estimates for the zero points (ZPs). Although the sky coverage of the SDSS spectra is smaller and sparser than \jplus photometry, it has the advantage that it can be used to calibrate those \jplus bandpasses that have no photometric counterpart. In particular, given the spectral coverage of SDSS spectra, these are used to calibrate the \jplus bandpasses from $J0395$ to $J0861$, including $g$, $r$ and $i$. With the installation of the BOSS spectrograph, the wavelength range of the spectra was extended to the blue, thus allowing the calibration of the $J0378$ band in those areas of the sky with BOSS spectra available. However, $u$ and $z$ bands fall out always of the covered range by SDSS spectroscopy. Given the large FoV of T80Cam at JAST/T80 (2\,deg$^2$), it is not rare having dozens of high quality SDSS stellar spectra in a single \jplus pointing. 

\item\textit{SDSS photometry.} The significant ($\sim$80\,\%) overlap between \jplus and SDSS footprints allows to calibrate the \jplus broad band observations against the corresponding ones in SDSS, after applying the needed color term corrections\footnote{Due to differences in the effective transmission curves between SDSS and \jplus photometric systems, it is needed to apply color term corrections to the SDSS photometry to obtain the corresponding \jplus photometry. These corrections are of particular importance in the case of the $u$ band where filters are known to be significantly different.}. This calibration technique is used to calibrate the $u$ and $z$ bands, uncovered by SDSS spectra.

\item\textit{Spectrophotometric standard stars}. The observation of spectrophotometric standard stars (hereafter SSSs) is the only procedure allowing to calibrate the full \jplus bandpass system, as long as the SSS spectra cover the full spectral range of the \jplus filter system. The main sources for the SSSs are the spectral libraries CALSPEC\footnote{\url{http://www.stsci.edu/hst/observatory/crds/calspec.html}}, the Next Generation Spectral Library\footnote{\url{https://archive.stsci.edu/prepds/stisngsl/}} and STELIB \citep{stelib}. Following the classical calibration procedure, each SSS is observed at different airmasses along the night to derive the extinction coefficient and the photometric ZP of the system. As it is well known, for this procedure to be accurate the atmospheric conditions must be very stable. The SSS technique is critical in the calibration of the $J0378$ filter, since neither SDSS photometry nor SDSS spectroscopy cover this bandpass.

\end{itemize}

The current procedures are thought to be applied to any single exposure or any combination of exposures in a given filter, independently of the observations in any other band. However, by combining the information from different bands, it should also be possible to apply methods that enable to anchor the calibration across the spectral range. One particularly promising approach is the use of the stellar locus \citep{covey2007,high2009,kelly2014}, which optimally suits in systems with large FoVs as those at the OAJ. This procedure profits from the way stars with different stellar parameters populate colour-colour diagrams, defining a well limited region (stellar locus) whose shape depends on the specific colors used. A specific stellar locus approach for the calibration of \jplus has been developed, obtaining consistent zero point calibrations over the full \jplus spectral range with $\sigma_{\rm ZP} \lesssim 0.02$. The details of this procedure and its application to \jplus data will be presented in a future work.

In Table~\ref{tab:J-PLUSmodes} we summarise the median zero points (ZPs) for each \jplus filter in the EDR. We also present the typical uncertainty in the calibration, ranging from $\sigma_{\rm ZP} \sim 0.06$ at the bluer bands to $\sim 0.04$ at the redder ones. These uncertainties were estimated from the comparison of duplicated sources in the \jplus overlapping areas, and should be accounted for in the photometry error budget of all sources. We also include the preliminary ZPs obtained through the use of stellar locus on the entire amount of \jplus data collected until November 2017 (rightmost column). A significant improvement in the ZP uncertainties is found in the latter case.

\begin{figure*}
\includegraphics[width=\textwidth]{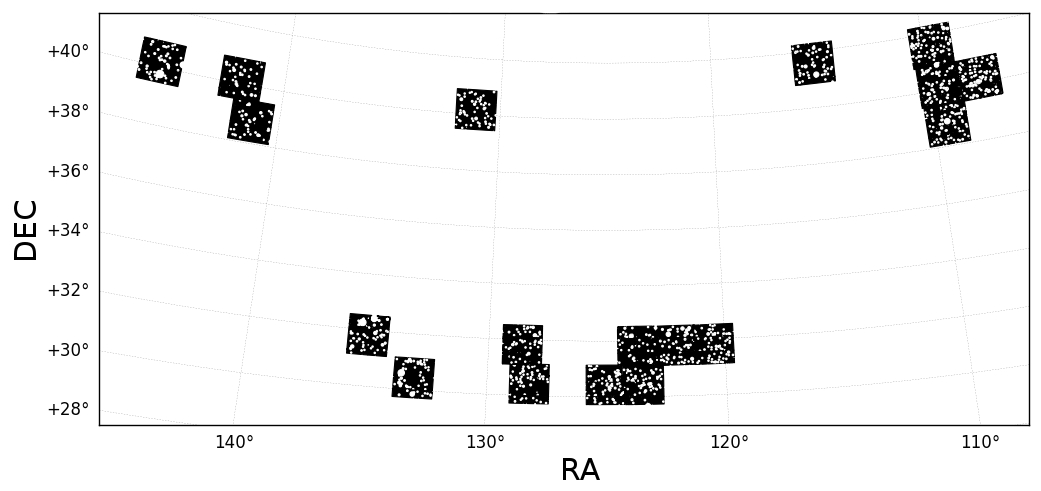}
\caption{Footprint of the \jplus EDR. It covers 31.7 deg$^2$ after masking low exposure regions, the surroundings of bright stars, observational reflections/artifacts, and overlapping areas.\label{fig:footprint_edr}}
\end{figure*}

\begin{figure*}
\includegraphics[width=0.5\textwidth]{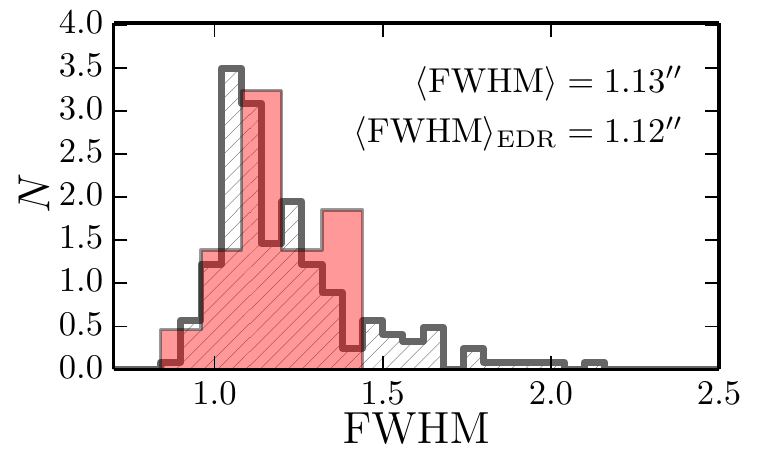}
\includegraphics[width=0.5\textwidth]{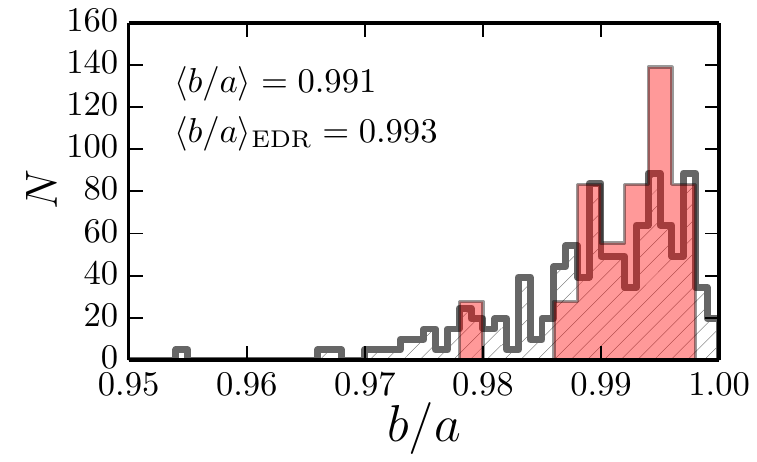}
\label{psf_stats}
\caption{FWHM and ellipticity ($b/a$ ratio) statistics of the \jplus EDR (red histograms) and the whole \jplus data gathered so far (gray hatched histograms) as measured in $r$ band on objects classified as stars. Histograms are normalised and thus represent probability. \label{fig:fwhm_edr}}
\end{figure*}

\section{\jplus Early Data Release (EDR)}
\label{sec:J-PLUS_EDR}

With the final \jplus strategy and depths frozen, we have identified and curated a subset of \jplus tiles that are representative of the \jplus project in terms of depth, PSF, photometric calibration accuracy, etc. These tiles compose the \jplus EDR and are publicly available at the \jplus web portal\footnote{\url{https://www.j-plus.es/datareleases/early_data_release}}.

The \jplus EDR comprises 18 \jplus pointings amounting to 31.7\,deg$^2$ after masking. The tiles belonging to the \jplus EDR were selected to i) have a limiting magnitude as close as possible to the final \jplus goal in all the filters simultaneously, ii) have no evident issues neither in the photometry nor in the source catalogues, and iii) define a reasonably compact region on the sky, as presented in Fig.~\ref{fig:footprint_edr}. 

The FWHM and ellipticity distributions of the \jplus EDR tiles in the $r$ band are shown in Fig.~\ref{fig:fwhm_edr}. These values are consistent with those of the parent \jplus data, and show average FWHM lower than $1.5$\,arcsec. The completeness for stars (point-like sources) and galaxies (extended sources) was computed by comparing with the SDSS dataset in the common areas (see \citealt{clsj17psmor}, for details). The median 50\,\% completeness in \jplus EDR is reached at $r = 21.5$ for stars and $r = 21.0$ for galaxies. For reference, the median 90\,\% completeness is reached at $r = 21.3$ for stars and $r = 20.5$ for galaxies.

\begin{figure*}
\begin{center}
\includegraphics[width=\textwidth]{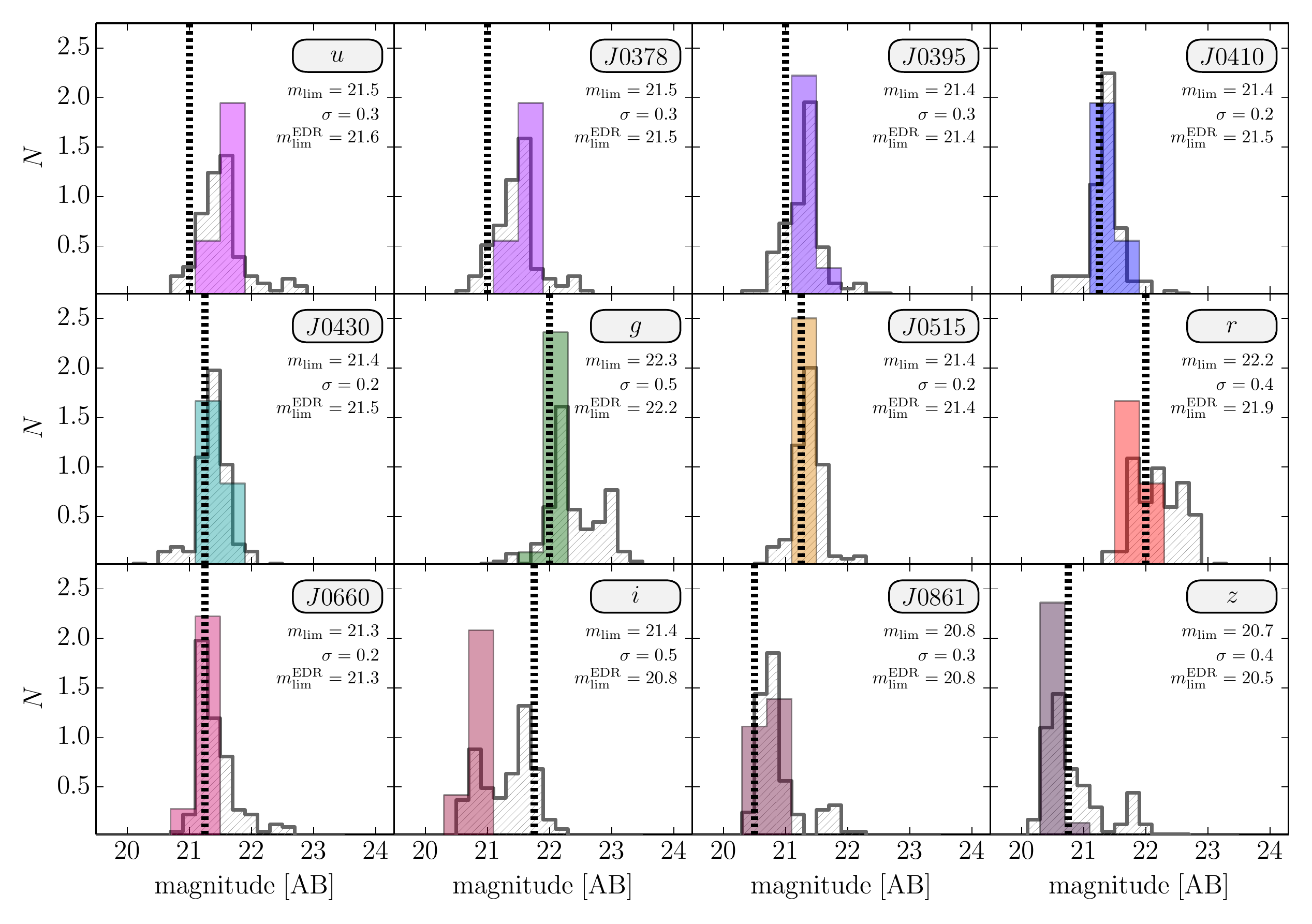}
\caption{Normalised distribution of the limiting magnitudes (3$\sigma$, $3^{\prime\prime}$ aperture) of
the \jplus EDR tiles (18 tiles, coloured solid histograms) and the whole \jplus data gathered so far (205 tiles, gray hached histograms). The black vertical lines mark the targeted \jplus limiting magnitudes as reported in Table~\ref{tab:J-PLUSmodes}. The legend in the panels provide the median of the limiting magnitudes in the two data samples. \label{fig:maglim_edr}}.
\end{center}
\end{figure*}

The EDR limiting magnitude distribution in the 12 \jplus bands is presented in Table~\ref{tab:J-PLUSmodes} and in Fig.~\ref{fig:maglim_edr}, where we indicate the target \jplus limits by means of a vertical dotted line. We can appreciate that, with the exception of the $i$ band, the EDR is fairly close to the targeted \jplus magnitudes. Additionally, as expected, the data is very homogeneous by construction, exhibiting variations in depth of $\sim 0.1$ magnitudes.

\begin{figure*}
\includegraphics[width=\textwidth]{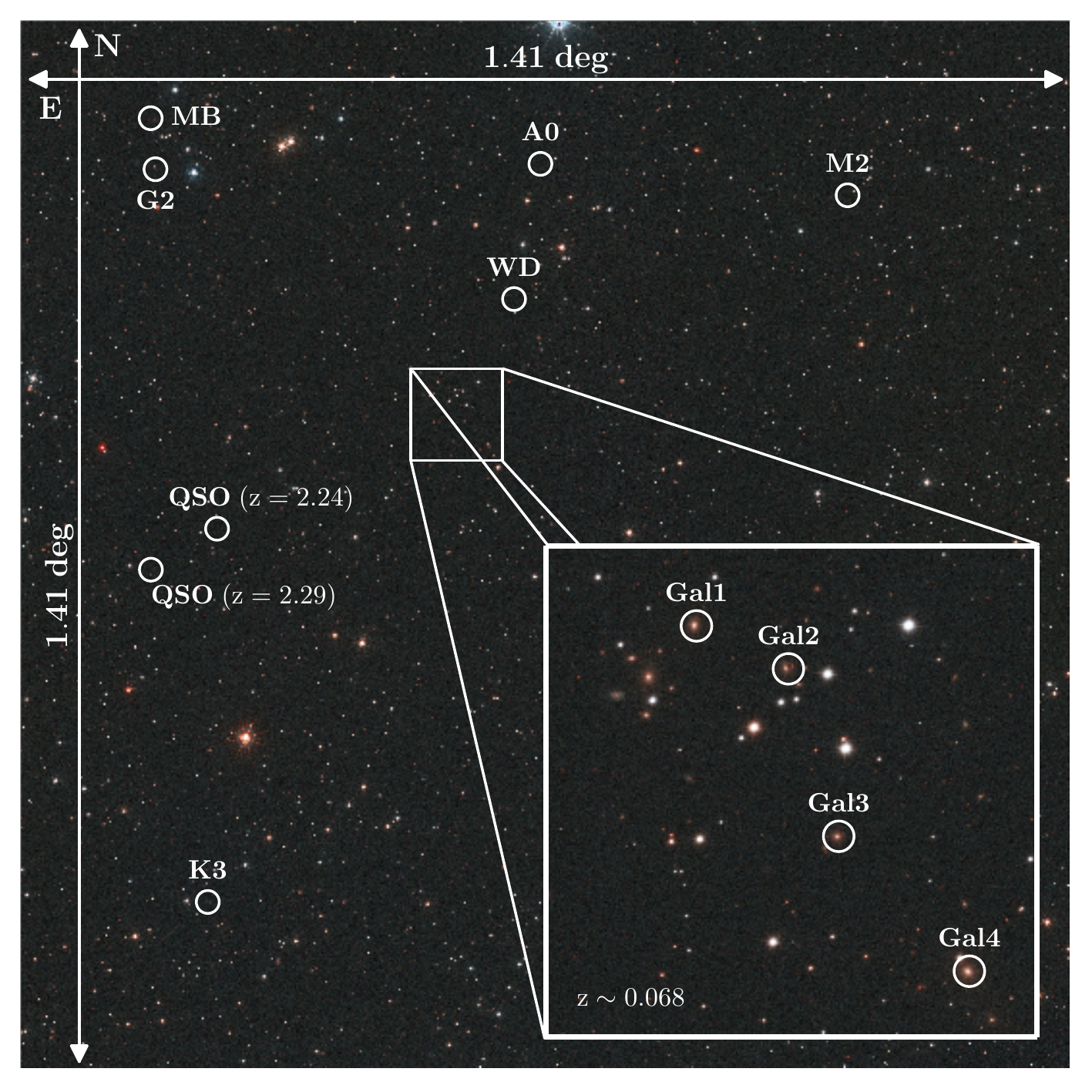}
\label{psf_stats}
\caption{Color composite of the \jplus EDR tile 4951. Several astrophysical objects analysed in the present paper are labelled in the figure: four MW stars of different spectral types (A0, G2, K3, M2); one white dwarf (WD), a minor body (MB) of the Solar System, four galaxies belonging to a $z = 0.068$ nearby cluster (Gal1, Gal2, Gal3, Gal4), and two high-$z$ quasars (QSOs).\label{fig:fig_edr}}
\end{figure*}

The EDR sources have been morphologically classified into stars or galaxies using a Bayesian approach \citep[see, for further details,][]{clsj17psmor}. In this work we find that \jplus sources exhibit two distinct populations in the magnitude vs. concentration plane, corresponding to compact (stars) and extended (galaxies) sources. We model the two-population distribution and used a number density prior based on EDR data to estimate the Bayesian probability of each source to be either a star or a galaxy. This procedure is applied in each pointing separately in order to account for varying observing conditions and the particular MW stellar density in that field. Finally, we combine the morphological information from $g$, $r$, and $i$ broad bands in an attempt to improve the classification of low signal-to-noise sources. The derived probabilities are used to compute the pointing-by-pointing number counts of stars and galaxies \citep{clsj17psmor}. The former increases as we approach to the Milky Way disk, and the latter are similar across the probed area. The comparison with SDSS in the common regions is satisfactory up to $r \sim 21$, with consistent number counts of stars and galaxies, and consistent distributions in concentration and $(g-i)$ colour spaces.

To summarise, the \jplus EDR tiles are representative of the parent \jplus data in terms of PSF and photometric quality, and they display depths similar to the \jplus data gathered so far. As a representative pointing, the color composite of the \jplus EDR tile 4951 is presented in Fig.~\ref{fig:fig_edr}. Likewise, Fig.~\ref{fig:ngc4470} provides the nearby galaxy NGC4470 as seen in the 12 \jplus bands.

\subsection{\jplus Science Verification Data (SVD)}\label{sect:SVD}
\begin{figure*}
  \centering
  \includegraphics[width=\textwidth]{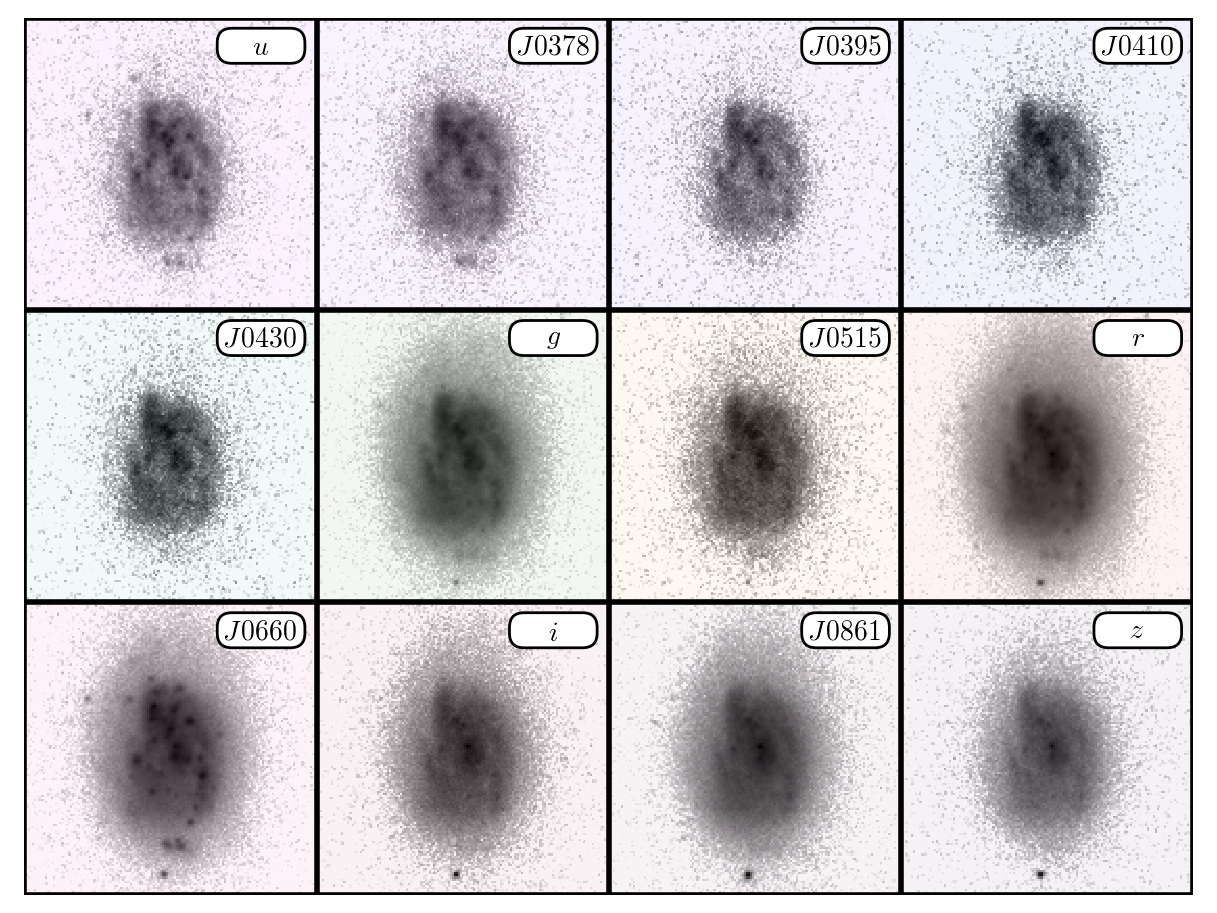}
  \caption{NGC4470 in the 12 \jplus bands. This nearby galaxy is ideal to illustrate how a typical HII galaxy is seen by \jplus, with plenty of young star-forming knots highlighted on the bluest filters ($u$ and $J0378$; covering [\ion{O}{ii}] emission line) and $J0660$ (covering H$\alpha$ and [\ion{N}{ii}] emission lines). As expected, these knots are essentially absent in the reddest bands, which rather trace the underlaying old stellar population of the galaxy.}
  \label{fig:ngc4470}
\end{figure*}

In addition to the \jplus EDR presented in the previous section, several tiles belonging to the \jplus science verification data (hereafter SVD) are also released with the present paper. These data were acquired before the beginning of \jplus, targeting particularly well-known objects to test and challenge the science capabilities of \jplus. The \jplus SVD pointings released with this paper\footnote{\url{https://www.j-plus.es/datareleases/svd}} are:

\begin{itemize}
\item 1 pointing centered at the MW globular cluster M15 \citep{bonatto17}.
\item 3 pointings covering continuously the galaxy clusters A2589 ($z=0.0414$) and A2593 ($z=0.0440$). The analysis of these data is presented in \citet{molino17}.
\item 3 pointings for the galaxies M101, M49 and the Arp313 triplet of galaxies. The nearby galaxies in these pointings, including NGC4470 ($z= 0.0079$, Fig.~\ref{fig:ngc4470}), are studied in \citet{logronho17} and \citet{sanroman18}.
\item 1 pointing centered at the Coma cluster of galaxies.
\item 6 pointings targeting several Milky Way planetary nebulae.

\end{itemize}

We stress that the \jplus SVD are not strictly considered \jplus data due to the following reasons: i) the photometric depths and observing conditions could be slightly different to the ones set for \jplus; ii) the observed fields can be either outside the \jplus footprint or have different coordinates than the predefined \jplus tiles; iii) a versioning of the SVD is not performed, i.e., the data of the different projects could have been processed with different versions of the pipeline; iv) the source catalogs are not necessarily the official catalogs created by the UPAD inside the \jplus collaboration, but could have been produced by different \jplus members; and v) the database back-end is neither available nor designed for these data. 

Still, the \jplus SVD provides very interesting data that deserve to be presented and made public to illustrate some science cases that can be developed with \jplus. Overall, the combination of \jplus EDR and SVD provides a total amount of 32 JAST/T80Cam pointings observed with the 12 \jplus filters, covering an area of $\sim 64$\,deg$^{2}$ before masking corrections. On the basis of these data, in the following section we focus on some of the most relevant \jplus science cases. 

\section{\jplus Science Cases}
\label{sec:J-PLUS_science}
The \jplus coverage of the optical SED with 12 photometric bands permits a great variety of studies, including MW stars (Sect.~\ref{sec:mw}), both nearby and much more distant galaxies at specific redshift windows (Sects.~\ref{sec:gal}, \ref{sec:zwin}), and variable objects (Sect.~\ref{sec:var}). In the next subsections we present representative examples of some of those topics, demonstrating the capabilities and the potential of \jplus.

\subsection{Exploring the MW halo}\label{sec:mw}

In the next years, the Gaia mission is going to provide the most detailed view ever made of the 3D-structure of the MW, improving our knowledge on its composition, formation, and evolution. Because of its survey strategy and performance, \jplus can help to complement Gaia's science on the MW halo. For instance, \jplus expands the Gaia wavelength coverage to both the near-IR and the UV. This extension in the UV range is of particular interest since the Gaia spectrophotometric system is not very efficient below 400\,nm. In this sense, the bluest \jplus filters can help to disentangle the determinations of extinction ($A_{\rm v}$) and effective temperature ($T_{\rm eff}$) by Gaia at fainter magnitudes and at the low metallicity regime. In addition, since there exists a time baseline of more than 20 years between the Tycho-2 and SDSS catalogues on one side, and \jplus on another, and the \jplus astrometry will be anchored on that of Gaia, \jplus can extend the proper motion determinations by Gaia down to a fainter limit of $g\sim22$\,mag. 

In general terms, the halo of the MW provides a unique laboratory to explore the nature of galaxy assembly in exquisite detail, the relation between the globular clusters and field star populations, the formation and evolution of the first generations of stars, and the nucleosynthesis in the early Universe, to name a few topics. Below we summarize a few examples showing the potential of \jplus on these fronts.
 
\subsubsection{Stellar parameters of MW stars}

\begin{figure*}[t]
	\centering
	\resizebox{0.49\hsize}{!}{\includegraphics{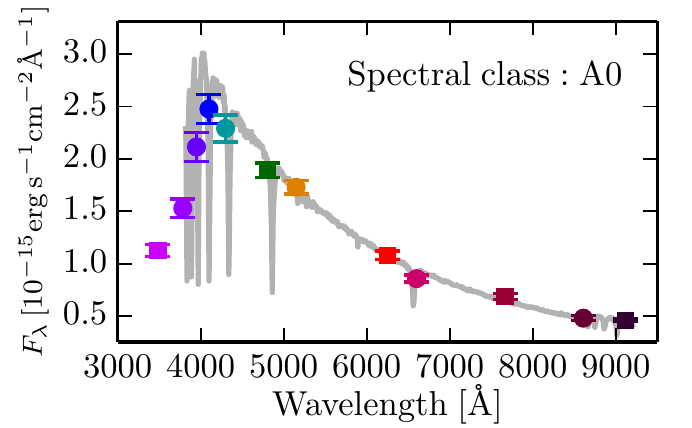}}
	\resizebox{0.49\hsize}{!}{\includegraphics{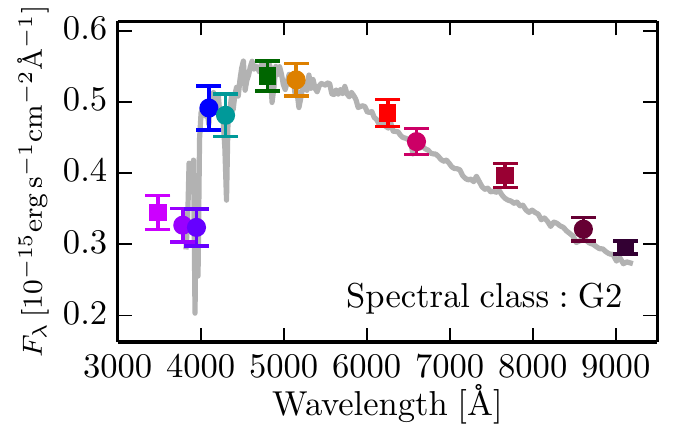}}\\
	\resizebox{0.49\hsize}{!}{\includegraphics{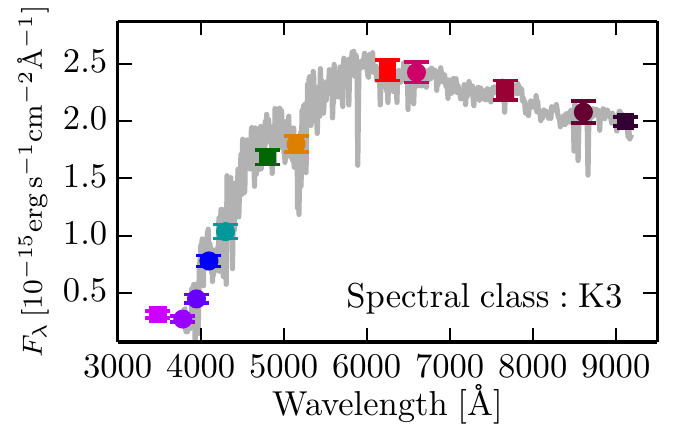}}
	\resizebox{0.49\hsize}{!}{\includegraphics{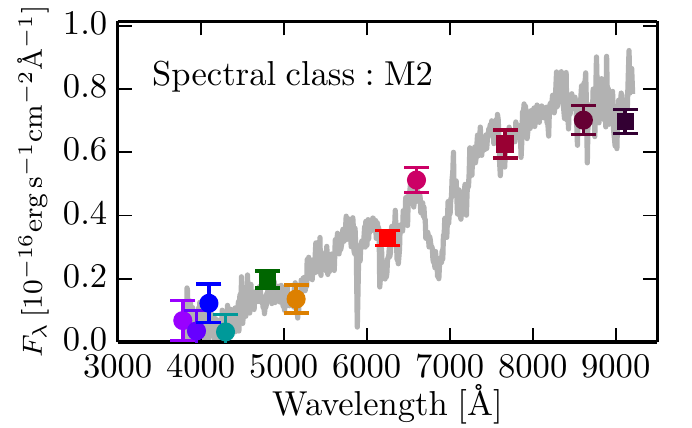}}
	\caption{\jplus photo-spectra of the four stars marked in Fig.~\ref{fig:fig_edr}. The gray lines show the SDSS spectra of these stars.}
	\label{fig:starclass}
\end{figure*}

\begin{figure*}[ht]
\centering
\includegraphics[trim={0cm 0.00cm 0 0.05cm},clip, scale=0.72]{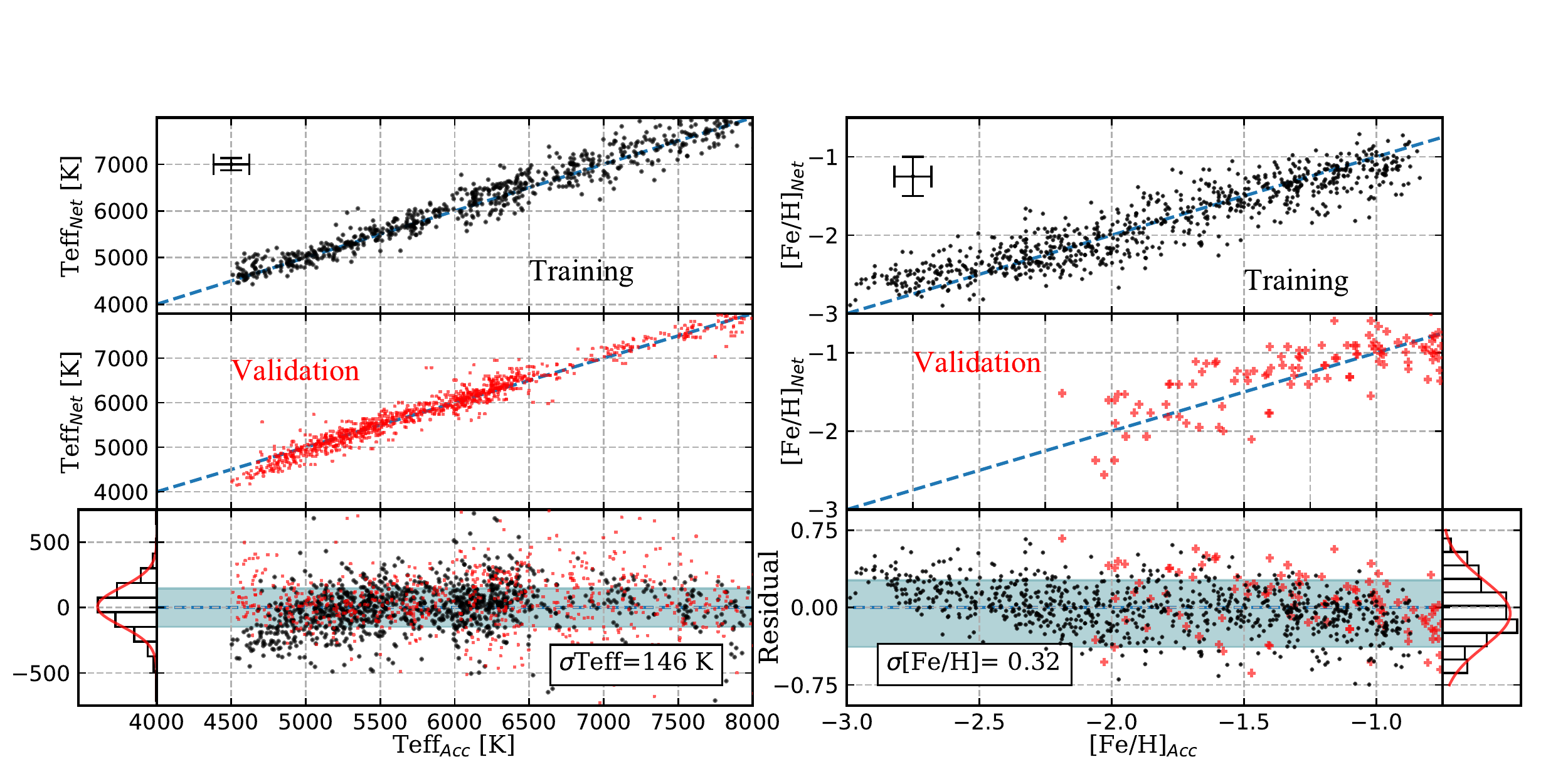}
\caption{  {\it Left panel:} Surface temperature predictions for a subset of \jplus EDR. {\it Right panel:} Metallicity predictions for a subset of \jplus EDR. Both networks were trained synthetically, the results of which denoted by black dots. The networks were then tested on EDR photometric inputs, shown in red. Estimates are compared in both cases to accepted values from the SSPP. Maximum-likelihood gaussian fits, shown in red, were produced for the residuals of the EDR photometric estimates.
}
\label{fig:teff_feh_plot}
\end{figure*}

With the advent of narrow-band photometric surveys such as \jplus it
is possible to provide a method to obtain the photometric stellar
parameter estimates required by the next generation of stellar
population studies. The narrow- and intermediate-band filters
implemented by \jplus are centred on key stellar absorption features,
which are sensitive to stellar parameters and chemical abundances,
including effective temperature ($T_{\rm eff}$), surface gravity
($\log g$), metallicity ([Fe/H]), as well as carbon abundance ([C/Fe])
and magnesium abundance ([Mg/Fe]).  Examples include the Ca\,{\sc ii}
H \& K lines, the CH $G$-band molecular feature close to H$\delta$,
and the Mg triplet, associated with the $J0395$, $J0430$, and $J0515$
filters, respectively. An illustration of the SEDs of different stellar spectral types as seen by \jplus is presented in Fig.~\ref{fig:starclass}.

While the \jplus filters are optimally placed along the SED to enable
detection of key absorption features, mapping their behaviour to overall
estimates of temperature, metallicity, and the elemental abundances of C
and Mg presents a highly degenerate problem. For example, at higher
temperatures, ($T_{\rm eff} \gtrsim 6000$\,K), the CH molecule
dissociates while the H$\delta$ line broadens to the extent of
disrupting the carbon feature. The Ca\,{\sc ii} K line provides an ideal
indicator for stellar metallicity, but also exhibits a temperature
dependence that prevents the direct use of line widths to estimate
metallicity.

Multiply-degenerate parameter spaces such as this can be approached
through the application of machine learning algorithms. Recently, a
subset of machine learning known as Artificial Neural Networks (ANNs)
have shown great promise as robust tools for stellar spectral classification
(\citealt{Kheirdastan:2016}). We have made use of these pattern
recognition tools for stellar parameter determination using the \jplus
photometric system.

The training dataset for the ANNs is taken from the SEGUE
\citep{Yanny:2009} database, in which stellar parameters were derived from the observed (flux
calibrated) spectra using the SEGUE Stellar Parameter Pipeline (SSPP,
\citealt{Lee:2008}). Synthetic magnitudes are first generated for these
spectra by convolution with \jplus filter response functions. Upon
verification of the proper network inputs, functional forms for these
synthetic magnitudes are then calibrated to the \jplus photometric
system using $\sim 1\,000$ overlapping observations from the \jplus EDR.

After training and for the final validation of the ANN methodology,
the temperature and metallicity networks are next applied to \jplus
photometry using the EDR. Figure
\ref{fig:teff_feh_plot} depicts the one-to-one fit and residual plot for
the network prediction against accepted SSPP estimate. The effective stellar temperature is recovered with a dispersion of $\sigma_{T_{\rm eff}} = 158$ K, while the expectations for the metallicity is $\sigma_{\rm [Fe/H]} = 0.31$ dex. We note deviations in the low temperature regime of the network estimates (T$_{\rm eff} < 4750$\,K). We attribute this underestimation to edge effects inherent in network estimates approaching the interpolation range of the training set in addition to an increasing uncertainty in the temperature estimates made by the SSPP in this range. 
An increase in scatter can be seen near T$_{\rm eff} = 6500$\,K, where we expect a presence of variable stars whose spectroscopic and photometric temperature were estimates determined at different phases in the stellar pulsation.

\begin{figure}[ht]
\centering
\includegraphics[width=0.5\textwidth]{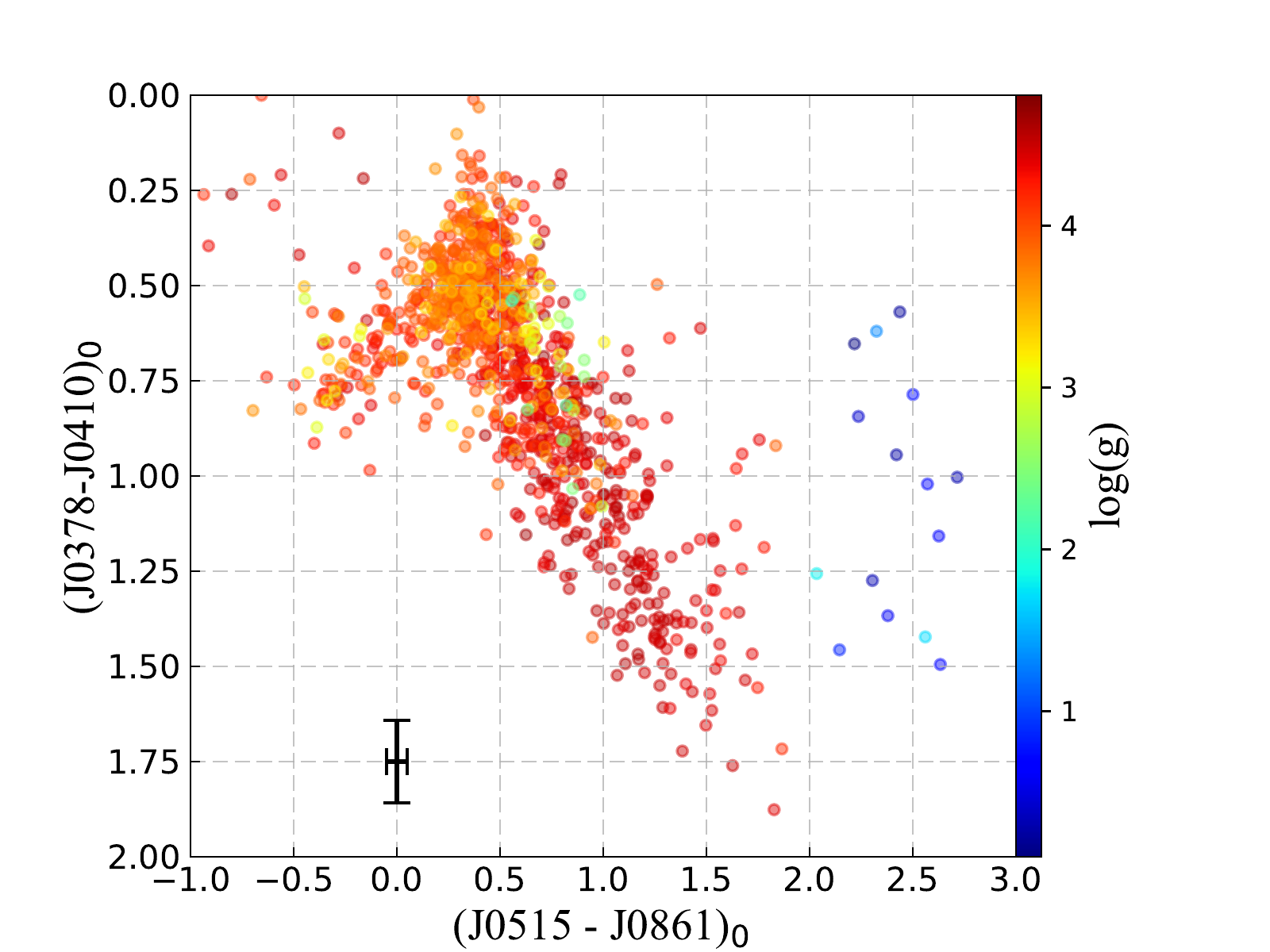}
\caption{Log $g$ separation in a surface gravity sensitive \jplus color-color plot.}
\label{fig:logg_separation.pdf}
\end{figure}

\jplus photometry also exhibits a crucial sensitivity to changes in
the surface gravity of stellar atmospheres. In particular, gravity
sensitive regions around the Balmer break (3700$\sim$4000\,\AA), as
well as the Paschen break, Ca{\sc ii} triplet lines and TiO bands
\citep[e.g.,][]{Cenarro2001a,2009MNRAS.396.1895C},
present in the $J0378$ and $J0861$ filters, respectively, can be
used. Pressure broadening affects the continuum in these regions,
corresponding to a detectable difference in flux between low and high
gravity environments \citep{Vickers:2012}. This behaviour is shown in
Fig.~\ref{fig:logg_separation.pdf}, where the gravity sensitive color
($J0378$-$J0410$)$_0$ is shown against the temperature dependent color
($J0515$-$J0861$)$_0$, and where the $0$ subscripts denotes ``reddening corrected".

\begin{figure}[ht]
\centering
\includegraphics[width=0.5\textwidth]{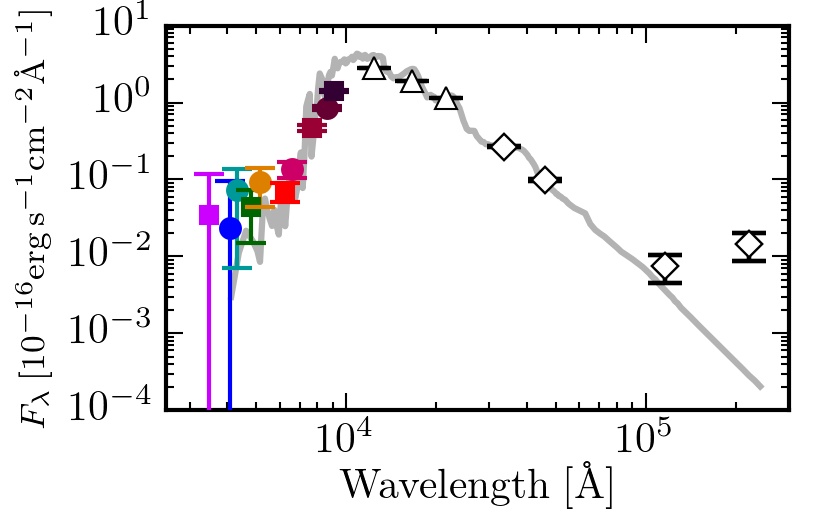}
\caption{\jplus photo-spectrum of a confirmed ultracool dwarf in the EDR footprint. \jplus measurements (coloured symbols) are complemented by 2MASS (white triangles) and WISE (white diamonds) data. The (smoothed) best fit model is provided by the grey line.}
\label{fig:JPLUS_jspec_UCD.jpg}
\end{figure} 

An approach alternative to ANNs is a direct search for optimal
atmospheric parameters via an interpolation in a model library,
obtained from Kurucz model atmospheres, as described
in \citet{allende16,allende18}. Our analysis is based upon the 
FERRE code \citep[][ and subsequent updates]{allende06}. The optimization
algorithm adopted in this case is the Boender-Timmer-Rinnoy Kan global
algorithm \citep{boender1982}.  The model fluxes considers
variations in effective temperature ($T_{\rm eff}$), surface gravity
($\log g$) and metallicity ([Fe/H]). We
perform tests using observations from the STIS Next Generation
Spectral Library \citep[NGSL;][]{gregg06,heaplindler07}\footnote{See also the online document \url{https://archive.stsci.edu/pub/hlsp/stisngsl/aaareadme.pdf}}.
After synthesizing \jplus photometry using the NGSL spectra and the
filter responses, we attempt to recover the atmospheric
parameters using the model grids and fitting techniques mentioned
above.  We focus our tests on the domain of the GK spectral types
(3500--6000\,K). Even under the assumption that the measurements of the \jplus filter responses are perfect, these tests should include systematic errors
associated with the imperfections in the model atmospheres and
spectral modelling.

The effective signal-to-noise per pixel of the observations in the
library, most likely dominated for many stars by systematic errors
(e.g., slit centering corrections), is expected to be in the range
20--50.  When the fluxes are integrated over the \jplus' passbands for the
the signal-to-noise is increased by a large factor.
The resulting, typical uncertainties in the recovery of the stellar parameters are
135~K in $T_{\rm eff}$, 0.6~dex in $\log g$,
and 0.4~dex in [Fe/H].  The stellar parameters of the stars in the
library are not uniformly distributed like in the tests, but they are
not heavily biased. Barring errors in the filter responses and our
ability to correct for atmospheric extinction, we conclude that at
least for signal-to-noise ratios lower than 50, i.e., 0.02~mag in the
\jplus photometry, random errors will likely dominate the error
budget. Note that these uncertainties are very close to those obtained when adopting the approach based upon ANNs.

\begin{figure}[t]
	\centering
	\resizebox{\hsize}{!}{\includegraphics{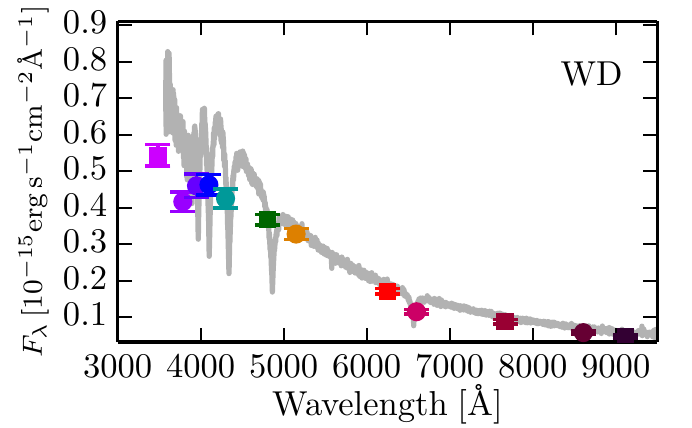}}
	\caption{\jplus photo-spectra of the white dwarf marked in Fig.~\ref{fig:fig_edr}. The gray lines show the BOSS spectra of this object.}
	\label{fig:wd}
\end{figure}

Thanks to the stellar temperature, metallicity, and gravity derived from \jplus photometry, a number of topics can be explored:

\begin{itemize}
\item {\it Search for metal poor stars}. With the exception of mass-transfer binaries or stars in advanced stages of evolution, long-lived low-mass, main sequence stars, in the absence of mass transfer interactions, retain in their atmospheres the chemical signature of their natal environments. As a result, the elemental abundances of ancient stars in the Galaxy provide the means to study the chemical evolution of the early Universe. The identification and subsequent detailed spectroscopic analysis of these old stars can provide crucial observational constraints for Galactic chemical-evolution models (\citealt{Salvadori:2010}), as well as placing fundamental constraints on the first mass function.

Obtaining spectroscopic estimates of metallicity ([Fe/H]) is a costly endeavour, however, requiring pre-selection and follow-up medium-resolution spectroscopic analysis \citep{Beers:2005}. While there now exist tens of thousands of stars with well-measured metallicities below 1/100th of the solar value ([Fe/H] $ < -2.0$), the numbers decrease rapidly with declining metallicity -- less than 25 ultra metal-poor stars (UMP; [Fe/H] $ < -4.0$) are known to date (\citealt{Placco:2015a}). 

We have demonstrated that the \jplus photometric system permits detection of
metal-poor stars across a wide range of temperatures, 4000--7000~K.
When using ANNs, this range can be further extended by broadening the distribution of
stars used to train the network, whether by making use of synthetic
spectra or empirical template libraries (\citealt{Kesseli:2017}).
As photometry is obtained for lower metallicity stars, calibration of estimates in the increasingly metal-poor regime ([Fe/H] < 2.0) will become possible using the methods described. It is anticipated that estimates of metallicity down to [Fe/H]$\sim-3.5$ will be achieved, which would
represent a major improvement from previous broad-band photometric
methods, where SDSS photometry reached saturation at [Fe/H] $ < -2$ (e.g.,
\citealt{Ivezic:2008}), or [Fe/H] $< -2.7$ (\citealt{An:2015}).  

\item {\it Ultracool dwarfs}. Defined as dwarf stars with spectral types later than M6, they include both very low-mass stars and brown dwarfs. Three different approaches are being carried out to find ultracool dwarfs using \jplus data, which should provide around 1\,600--2\,400 new candidates by the end of the survey:

\begin{enumerate}
\item{Photometric search}. A preliminary discriminator for candidate objects is built upon the subsample of spectroscopically confirmed M dwarfs with known spectral types \citep[][]{west2011} that count with \jplus photometry. The (\jplus based) $i-z$ color of those sources is used to search for candidates in the entire \jplus database. Using a Virtual Observator (VO) tool like VOSA \citep{bayo2008} we are able to obtain SEDs for those candidates, and provide estimates for their effective temperature ($T_{\rm eff}$). Objects with effective temperatures corresponding to spectral types later than M6 \citep[according to][]{west2011} are flagged as candidate ultracool dwarfs.  Using {\tt Aladin} \citep{bonnarel2000} we are able to identify the objects previously reported in the literature. One of these is shown in Fig.~\ref{fig:JPLUS_jspec_UCD.jpg}.
\item{Kinematic search}. Given that ultracool dwarfs are intrinsically faint objects, they must be located in the solar vicinity, and thus should have, on average, a high relative motion. Using the information available in Simbad \citep{wenger2000} we defined a limit in proper motion of 50 mas/y to distinguish between
ultracool dwarfs and other types of objects with a high degree of
completeness and low degree of contamination. Effective temperatures
of the candidates ultracools will be subsequently computed using VOSA.
\item{Search based on Artificial Intelligence techniques}.  Using the
\jplus color combinations that best reproduce SED shapes, different
approaches will be used to identify new ultracool dwarfs.
\end{enumerate}

\item {\it Detection and characterization of white dwarfs}. The bluest \jplus bands permit to characterize the white dwarf (WD) population. As example, we present the \jplus photo-spectra of a spectroscopically confirmed WD in Fig.~\ref{fig:wd}. From the experience from the Sloan and the Gaia surveys, we expect that \jplus will characterise around 3 WDs per square degree, or about 30\,000 WDs for a final area of 8\,500\,deg$^2$.

\item {\it Search for stellar streams in the MW halo.} The efficiency of detecting stellar streams in the MW halo is heavily dependent on the possibilities to filter out candidate sub-populations from the background halo stars, which in general have very similar properties, also being old, low in mass, and poor in metals. While already the SDSS broad-band filters have allowed to search for and detect stellar streams using the Hess diagrams \citep{2006ApJ...643L..17G}, the stellar parameters based on the \jplus filter set are even more promising for such a work, especially in combination with the powerful but shallower dataset of stellar kinematics from the Gaia mission.

\end{itemize}

\subsubsection{Searching for planetary nebulae and symbiotic stars}

\begin{figure*}
	\centering
	\includegraphics[width=0.49\textwidth]{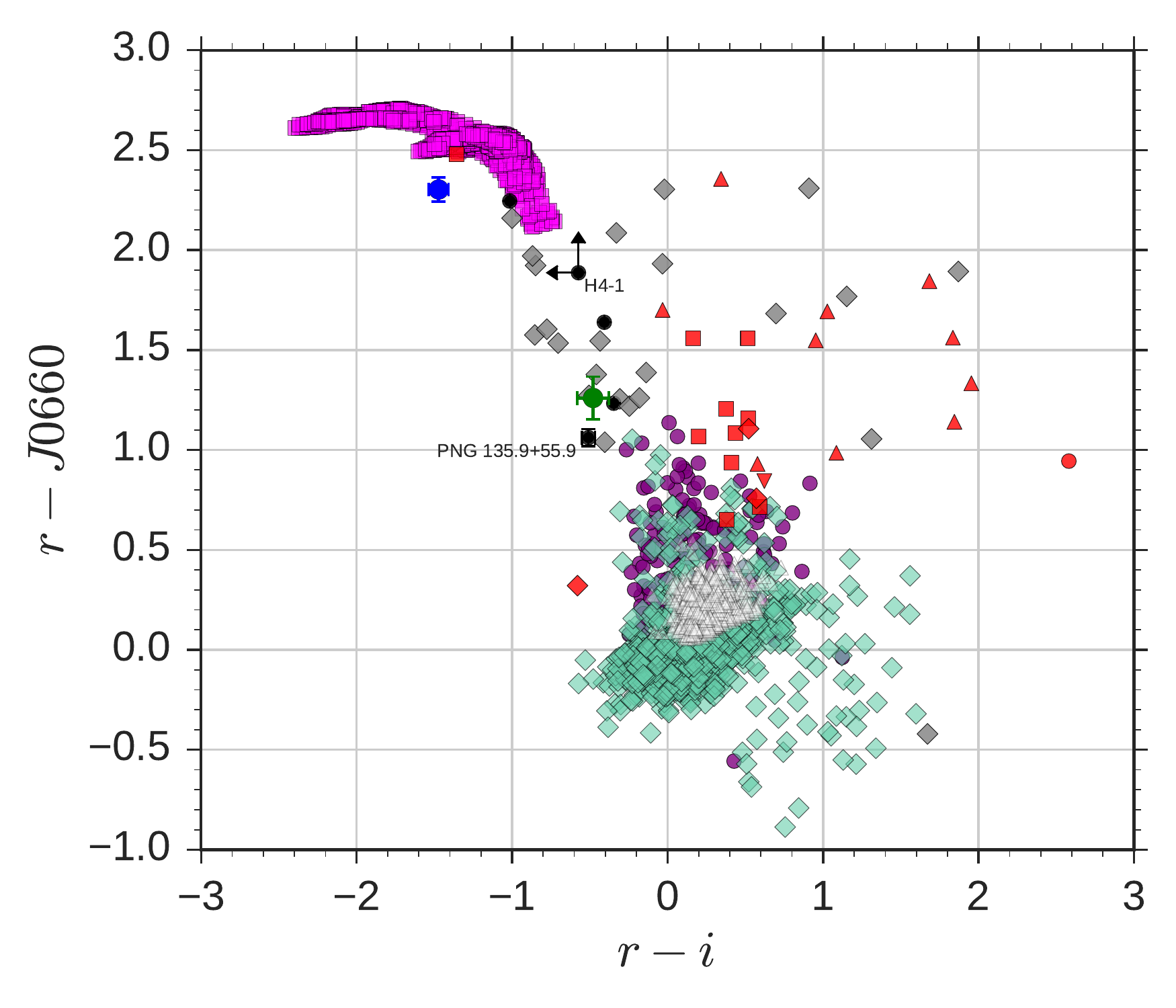}
	\includegraphics[width=0.49\textwidth]{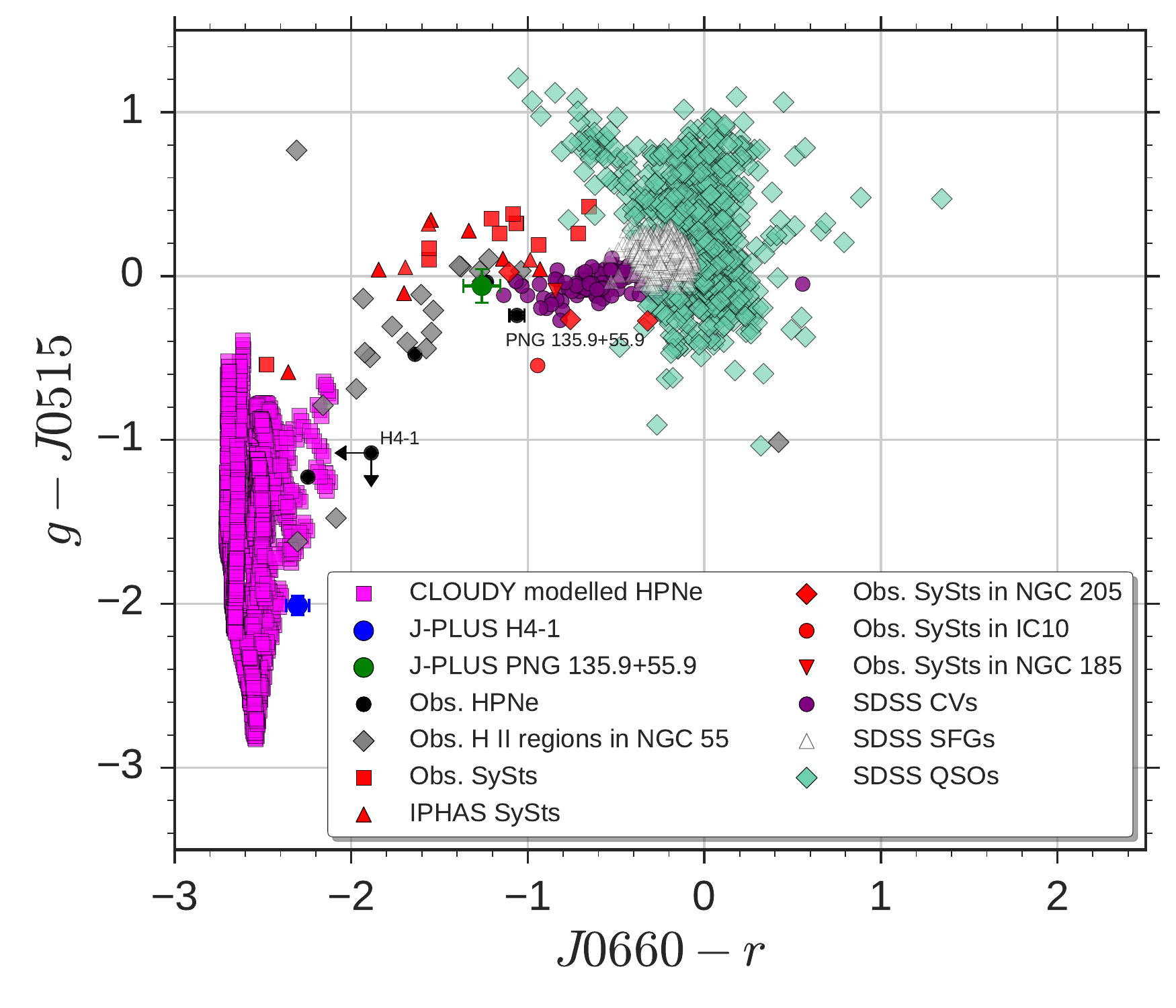}
\caption{{\it Left panel:} \jplus $r$-$J0660$ vs. $r-i$ colour-colour diagram, equivalent to IPHAS $r'-$H$\alpha$ vs. $r'-i'$. {\it Right panel:} $g-J0515$ vs. $J0660$ diagram. Blue and green symbols with errorbars are the \jplus observations for H~4-1 and PNG~135.9+55.9, respectively. Included in the diagrams are families of {\sc cloudy} modelled HPNe (pink boxes) spanning a range of HPNe properties. Black circles represent HPNe H~4-1 and PNG~135.9+55.9 spectra from SDSS, DdDM-1 \citep{1998ApJ...493..247K}, NGC~2022 \citep{2003PASP..115...80K}, and MWC~574 \citep{2007A&A...467.1249P}. Gray diamonds represent H~{\sc ii} regions in NGC~55 \citep{2017MNRAS.464..739M}. Red boxes display \citet{2002A&A...383..188M} SySts and red triangles correspond to IPHAS symbiotics \citep{2014A&A...567A..49R}. Red diamonds, circles and inverted triangles refer to the SySts in external galaxies, NGC~205 \citep{goncalves2015}, IC~10 \citep{goncalves2008} and NGC~185 \citep{goncalves2012}, respectively. Violet circles correspond to SDSS cataclismic variables (CVs). Black, empty triangles refer to SDSS star-forming galaxies, SDSS SFGs. SDSS QSOs at different redshift ranges (light blue diamonds) were selected, since some of their lines are mimics of the H$\alpha$ and/or [O~{\sc iii}] 500.7~nm emission in the local Universe. 
}
\label{fig:pne_cmds}
\end{figure*}

\begin{figure*}
	\centering
	\resizebox{0.49\hsize}{!}{\includegraphics{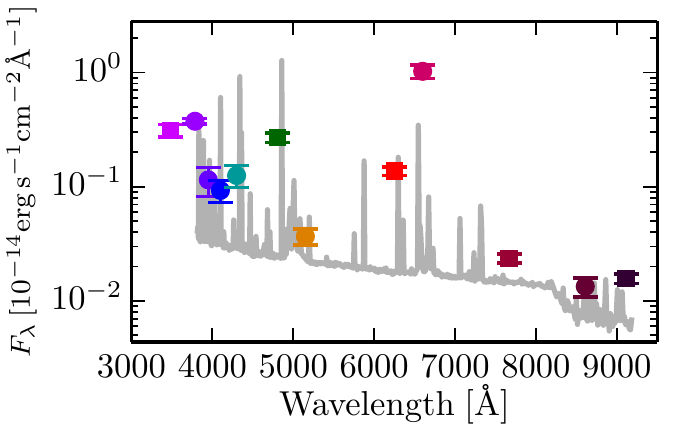}}
	\resizebox{0.49\hsize}{!}{\includegraphics{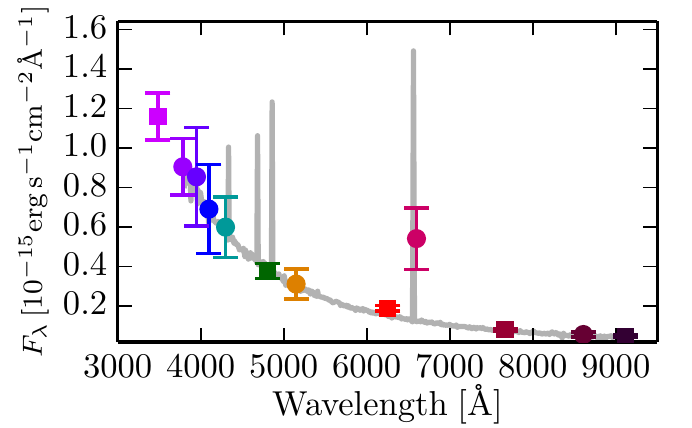}}
\caption{\jplus photo-spectra of the PNe H~4-1 ({\it left panel}) and PNG~135.9+55.9 ({\it right panel}). In both spectra the strong H$\alpha$ emission is easily perceptible.}
\label{fig:pne_jspectra}
\end{figure*}

Here we address the study of distinct families of systems, namely planetary nebulae (PNe) and symbiotic systems (SySts), the reason being that both involve the ionization of circumstellar gas, either ejected and ionized by a white dwarf itself (PNe), or associated to the stellar wind of a red giant that is ionized by high energy photons generated in the accretion onto a companion star (SySts).

PNe are essential objects to study the chemical evolution of galaxies \citep[see, for a review,][]{Magrini12} as well as to determine extragalactic distances \citep[see, for instance,][]{Ciardullo12}. On the other hand, the binary nature of the symbiotic systems make them one of the possible supernova Ia progenitors \citep{1973ApJ...186.1007W, 2010ApJ...719..474D, 2012Sci...337..942D}. Both enrich the interstellar medium, with PNe providing important information about the stellar nucleosynthesis of low- to intermediate-mass stars (0.8--8\,M$_{\odot}$). However, both families of objects are intrinsically different, with
SySts constituting interacting binary systems which can alter significantly the evolutionary and chemical path of the stars involved.  

A continuum marked by emission lines is the optical fingerprint of PNe while this is the historical definition and usual way to identify SySts. In fact, the PNe and the population of known SySts revealed in optical surveys usually present strong hydrogen and helium recombination lines (e.g., H$\alpha$, \ion{He}{ii}) as well as a number of low- and high-excitation lines (e.g., [\ion{O}{iii}], or even [\ion{Fe}{v}] and [\ion{O}{vi}] in the case of SySts). They are thus ideal targets to be searched for by \jplus in the direction of the north Galactic halo. Moreover, SySts' spectra (see, for instance, \citealt{2002A&A...383..188M}) are also rich in absorption features, like TiO, VO and others, due to the presence of the cool companion (with M, K or G spectral types).

Despite the strong efforts to increase the number of Galactic PNe presently known ($\sim$3000, \citealt{2012MNRAS.427.3016P}), only 14 of them are classified as halo PNe (HPNe; \citealt{2015ApJS..217...22O}). Nevertheless, more PNe are certainly hidden in the Galactic halo, since the mass range of PNe (SySts) dominates the halo stellar population. Regarding SySts, 251 are known in our Galaxy, and only 5\,\% of those are located in galactic latitudes of $|b|>30$\,deg, (Akras et al., submitted). 

Given the scarcity of PNe and SySts in the Galactic halo, we aim at performing the first systematic search of these evolved objects, at high latitudes, by using the 12 \jplus filters. These twelve measurements provide a number of colour combinations by far more numerous than those used in any previous search (for instance in IPHAS, \citealt{2005MNRAS.362..753D}; or VPHAS, \citealt{2014MNRAS.440.2036D}), making it much easier to distinguish PNe and SySts with strong emission lines from other emission line objects in the optical. In Fig.~\ref{fig:pne_cmds} we show that \jplus is able to optimally separate different families of emission line objects. In particular, we reproduce the equivalent IPHAS $r'$-H$\alpha$ vs. $r'$-$i'$ colour-colour diagram, which in IPHAS as well as in our survey can efficiently highlight and discriminate the HPNe locus with respect to their mimics. The IPHAS version of this diagram is found in Fig.~1 of \citet{2009A&A...504..291V}. Other combinations of colours have been explored for \jplus (Guti\'errez-Soto et al., in prep.). For instance, the $g$-$J0515$ vs. $J0660$-$r$ diagram, in Fig.~\ref{fig:pne_cmds}, is also found to provide a good PNe(SySts) discrimination. 

In order to build these diagrams, \jplus synthetic magnitudes were computed for different kinds of objects (as given in the caption and legend of Fig.~\ref{fig:pne_cmds}). These synthetic data were obtained by means of observed spectra and a grid of models for HPNe provided by the photo-ionization code {\sc cloudy} \citep{2013RMxAA..49..137F}. In addition, the \jplus SVD of two known HPNe (H~1-4 and PNG~135.9+55.9) are presented in Fig.~\ref{fig:pne_cmds}. This figure shows that the locus occupied by the two HPNe observed by \jplus SVD is compatible with that associated with the synthetic data. In the case of PNG~135.9+55.9, the uncertainties of the spectroscopic data constitute the lower limits for the errors associated to the intensities of the important lines within each filter \citep{2002A&A...395..929R}. In regard of H~1-4, we need to point out that the SDSS spectrum is saturated in [\ion{O}{iii}] 5007~\AA\, as well as in H$\alpha$. Hence, the synthetic \jplus' $J0660$, $r$, $J0515$ and $g$ magnitudes are either upper or lower limits, as indicated in the colour-colour diagrams. Moreover, we note  
that H~1-4 is found to lie very close to the regime where halo PNe are expected to be, while the PNG~135.9+55.9 lies rather farther away, in the regime of HII regions. 
This happens because the PNe in the diagrams have different excitation degrees, and those of lower-excitation are, photometrically, indistinguishable from H II regions. Only the inclusion HeII\,4686\,\AA magnitudes could avoid this problem. Fortunately, only a few Galactic H II regions will be detected by \jplus, whose observations will be towards the halo, while massive and young stars are typically in the disk. In Fig.~\ref{fig:pne_jspectra} we present the \jplus photo-spectra of these two HPNe, in which the H$\alpha$ emission line is easily perceptible. 

Machine learning algorithms such as linear discrimination analyses and classification trees, among others, will be applied to this work in order to find more general ways to discriminate PNe and SySts from other strong emission-line sources that will be observed by the \jplus survey.

As for the case of SySts candidates, additional clues can also be provided from observations available in X-rays surveys (e.g., XMM-Newton, Chandra, and Swift), especially for the cases displaying weak emission lines \citep{mukai2016} and that can represent a significant fraction of SySts that have remained invisible to optical surveys up to present days. 

\subsubsection{Galactic Globular Clusters: the case of M15}


The wide-field capabilities of JAST/T80Cam are an excellent tool to study a wide-range of integrated properties of Galactic globular clusters, such as mass and luminosity segregation, luminosity functions, and total mass, among others. In addition, the data can also be used to search for multiple populations\footnote{Multiple stellar populations have been detected in a large fraction of galactic globular clusters (see e.g. \citealt{Gratton2012}, \citealt{Bastian2015}).} in different regions of the clusters, as well as any large scale effects that may be related to them, such as disruption seen in the form of tidal tails. We expect to find around 10 globular clusters falling in the footprint of \jplus.

M\,15 (NGC\,7078) is an old ($\ga$10\,Gyr) and very metal-poor, [Fe/H]=-2.3, \citep[][]{Carretta2009}, globular cluster, located $\sim$10\,kpc away from the Sun and the Galactic centre \citep{Harris2010}. 
Interestingly, a previous Hubble Space Telescope (HST) analysis \citep{Larsen2015} indicated a split in the lower Red Giant Branch (RGB) stellar distribution, with first and subsequent generation populations dominating at different clustercentric distances. To illustrate this science case, M\,15 has been observed during \jplus science verification 
well beyond the tidal radius ($R_{\rm tidal} = 21.5$\,arcmin, \citealt{Harris2010}) with uniform photometry. The full analysis for this cluster using \jplus SVD is presented in \citet{bonatto17}.  


Because of crowding, \citet{bonatto17} employ {\tt IRAF/Daophot} to build photometric catalogs for the 12 bands of the whole M15 field. Calibration is performed with the zero-points provided by the UPAD. Here we aim to emphasize the potential of \jplus on the analysis of globular clusters, particularly with respect to detecting multiple stellar populations in the RGBs of M15. An excellent example is the colour-magnitude diagram (CMD) $J0378$ versus $(J0378-J0861)$ shown in Fig.~8 of \citet{bonatto17}, 
in which the RGB appears to split into two sequences. The splits are considerably wider than the ones expected from photometric scattering, as can be inferred from the average error bars. Fiducial lines, built to represent the mean path followed by each sequence, are included in that figure as a visual aid. Visually considering the sequence split and guided by the fiducial lines, \citet{bonatto17} separate the stars in blue and red sub-samples, and take their magnitudes in the other \jplus filters, noticing that the split vanishes when broad-band filters such as $g$ and $r$ are considered. This suggests a relation to light-element abundance differences, since $J0378$ and $J0861$ are sensitive to N and Ca abundances. See \citet{bonatto17} for full details on this particular science case.


\begin{figure*}
	\centering
	\resizebox{0.49\hsize}{!}{\includegraphics{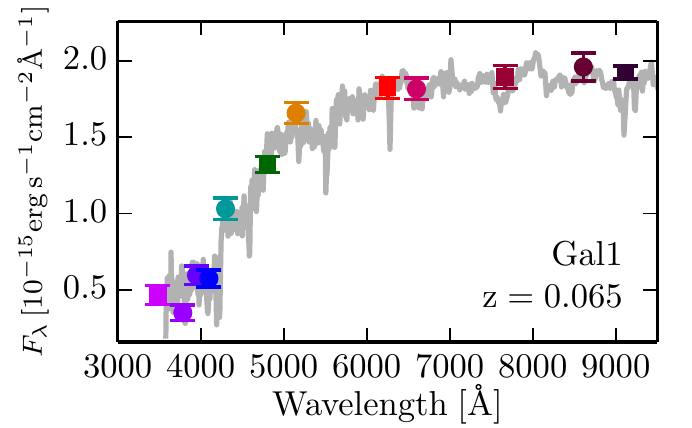}}
	\resizebox{0.49\hsize}{!}{\includegraphics{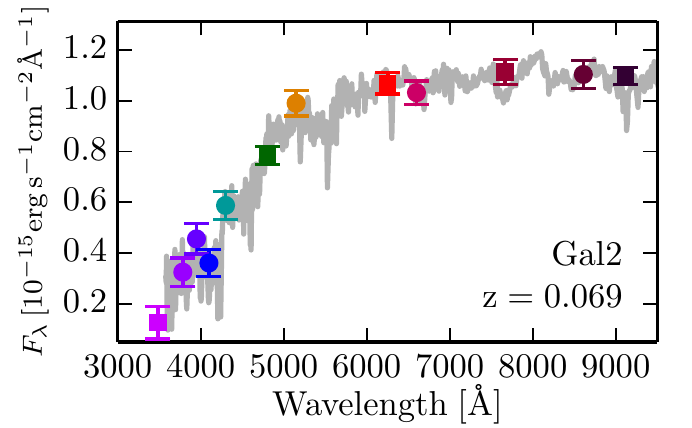}}\\
	\resizebox{0.49\hsize}{!}{\includegraphics{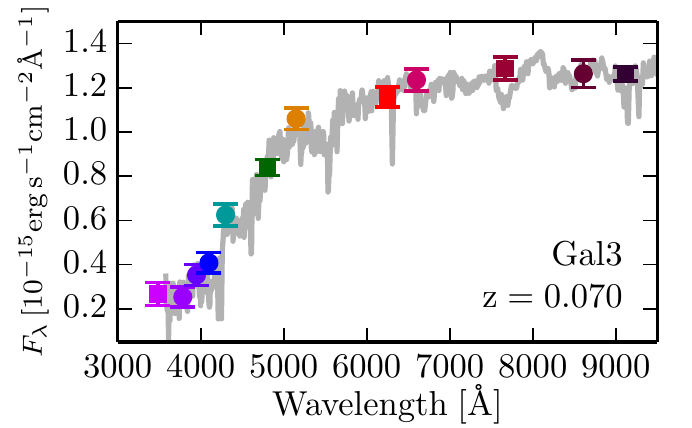}}
	\resizebox{0.49\hsize}{!}{\includegraphics{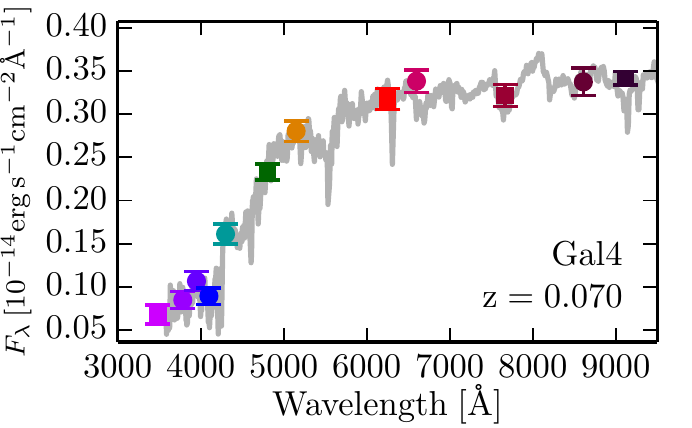}}
	\caption{\jplus photo-spectrum of the four galaxies marked in Fig.~\ref{fig:fig_edr}. The SDSS spectra, scaled to match the $r$ band total magnitude from \jplus, are presented in grey.}
	\label{fig:ell_photoz}
\end{figure*}

\begin{figure}
\centering
\includegraphics[width=0.5\textwidth]{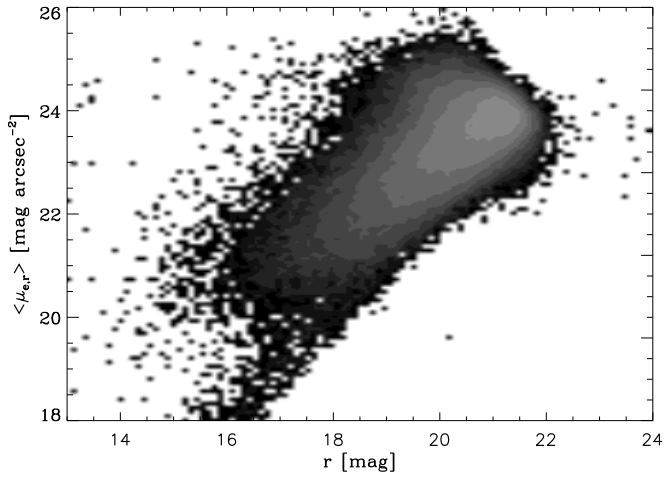}
\caption{Location of EDR galaxies in the plane defined by mean effective surface brightness $\langle \mu_{e,r}\rangle $ and apparent magnitude, both in the $r$ band.}
\label{fig:mu_er}
\end{figure}

\subsection{Studies in the nearby Universe}\label{sec:gal}
Large spectroscopic surveys of galaxies fail to map, at low redshifts, the whole extent of spatially resolved galaxies and their close environment. To solve the first issue, systematic studies based on integrated field units (IFUs) have been conducted (e.g., the Sauron project, \citet[][]{sauron}; ATLAS$^{\rm 3D}$, \citealt{ATLAS3D2011}; CALIFA, \citealt{CALIFA2012}; SAMI, \citealt{SAMI2012}; MaNGA, \citealt{manga}). With these, one can analyze galaxies as a whole and look for gradients or trends in specific properties. However, while spectroscopy is usually less efficient and more time consuming than photometric studies, IFU studies tend to be limited to bright and size-selected objects.

Half-way between classical photometry and spectroscopy, \jplus will build a formidable legacy data set by delivering low resolution spectroscopy for every pixel over a large, contiguous area of the sky. This IFU-like character, allowing a 2D pixel-by-pixel investigation of extended galaxies, will lead to much larger galaxy samples than classical multi-object spectroscopic surveys. In addition, no sample selection criteria other than the photometric depth in the detection band will result in a uniform and non-biased spatial sampling, thus allowing for environmental studies. Furthermore, provided that direct imaging is more efficient than spectroscopy, \jplus will be generally deeper than traditional spectroscopic studies, enabling better access to galaxy outskirts and fainter companions.
 
Figure~\ref{fig:mu_er} shows the limits of \jplus for studies in the nearby Universe. This figure represents the location of the EDR galaxies (MORPH\_PROB\_STAR\footnote{This flag is found in the database table {\tt jplus.stargalclass} and provides the probability of an object for being a star based upon morphological arguments. See \citet{clsj17psmor} for further details.} $<0.2$) in the plane defined by the mean effective surface brightness $\langle \mu_{e,r} \rangle$ and the apparent magnitude $r$, both in the $r$-band. Note that down to $r=20.5$ \jplus can detect galaxies brighter than $\langle \mu_{e,r}\rangle = 25.0$. In addition, objects down to $r=22.0$ have been observed by \jplus with $\langle \mu_{e.r}\rangle$ brighter than $\approx 23.5$. This depth makes to \jplus an ideal survey to select targets for spectroscopic follow-ups by future surveys such as WEAVE\footnote{\url{http://www.ing.iac.es/weave/}}, MANGA\footnote{\url{http://www.sdss.org/surveys/manga/} }, or DESI\footnote{\url{http://desi.lbl.gov}}, among others. 

\jplus will also constitute a reference survey in the optical, due to its unique color information, in cross-correlation studies with surveys in other wavelengths, such as the IR, X-rays or $\gamma$ rays. The study of energetic sources by missions like XMM\footnote{\url{http://sci.esa.int/xmm-newton/}}, Chandra\footnote{\url{http://chandra.harvard.edu/about/axaf\_mission.html}}, Swift\footnote{\url{https://swift.gsfc.nasa.gov}}, Suzaku\footnote{\url{https://www.nasa.gov/content/suzaku-mission-overview}} or Fermi\footnote{\url{https://fermi.gsfc.nasa.gov}} will be complemented by the multi-color information provided by \jplus, which should be of particular relevance in the local universe. 

In this section, we demonstrate \jplus capabilities to study the properties of nearby galaxies, focusing on their photometric redshift (Sect~\ref{sec:photoz}), 2D star formation rate (hereafter SFR, Sect~\ref{sec:2dsfr}), 2D stellar content (Sect~\ref{sec:2dsps}), and environment (pairs, groups, and clusters; Sect~\ref{sec:2denv}).

\begin{figure}
\includegraphics[width=9.0cm]{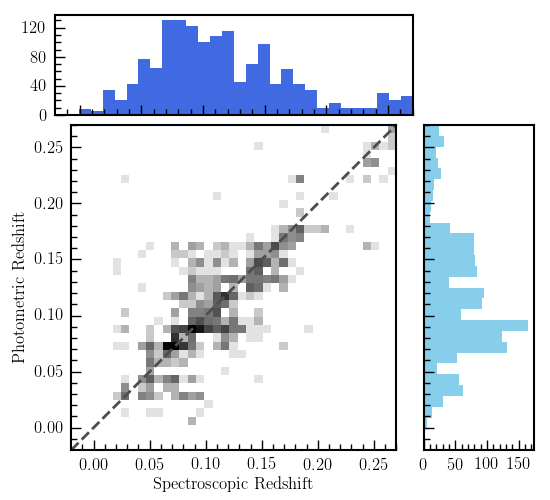}
\caption{\jplus photometric redshifts vs SDSS spectroscopic redshifts for the common sources in the \jplus EDR. The side panels show the projection in redshift space of the photometric ({\it right}) and spectroscopic ({\it upper}) values.}
\label{fig:photo_vs_spec}
\end{figure}

\begin{figure}
\includegraphics[width=9.0cm]{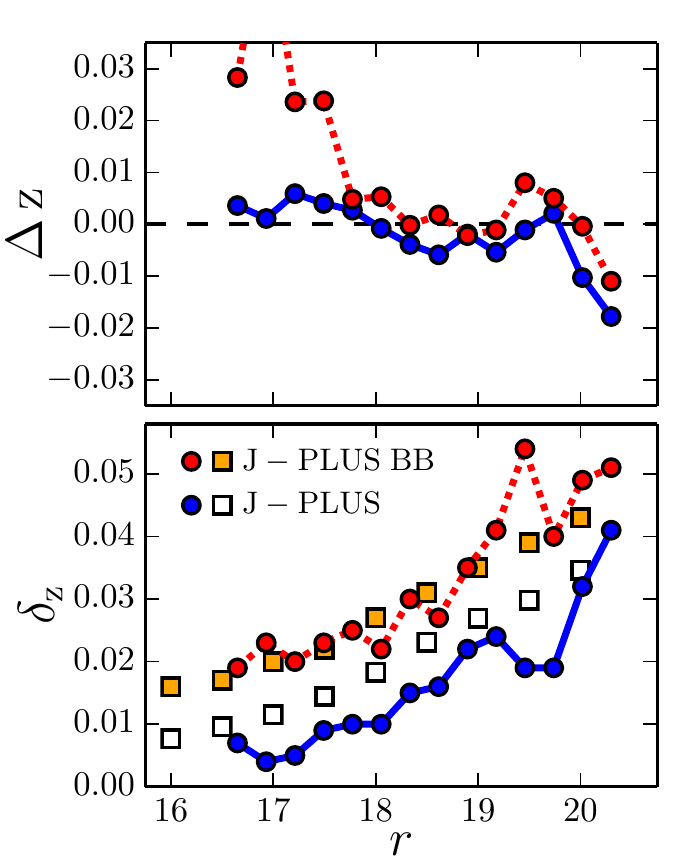}
\caption{\jplus photometric redshift bias ({\it top panel}) and precision ({\it bottom panel}), estimated from EDR data. The blue circles show the results with the twelve \jplus filters, and the red circles with only the $ugriz$ broad bands used. The orange (five bands) and white (twelve bands) squares show the expected photo-$z$ precision from the realistic simulations in \citealt{molino17}.}
\label{fig:photozsimul}
\end{figure}

\begin{figure*}
  \centering
  \includegraphics[width=\textwidth]{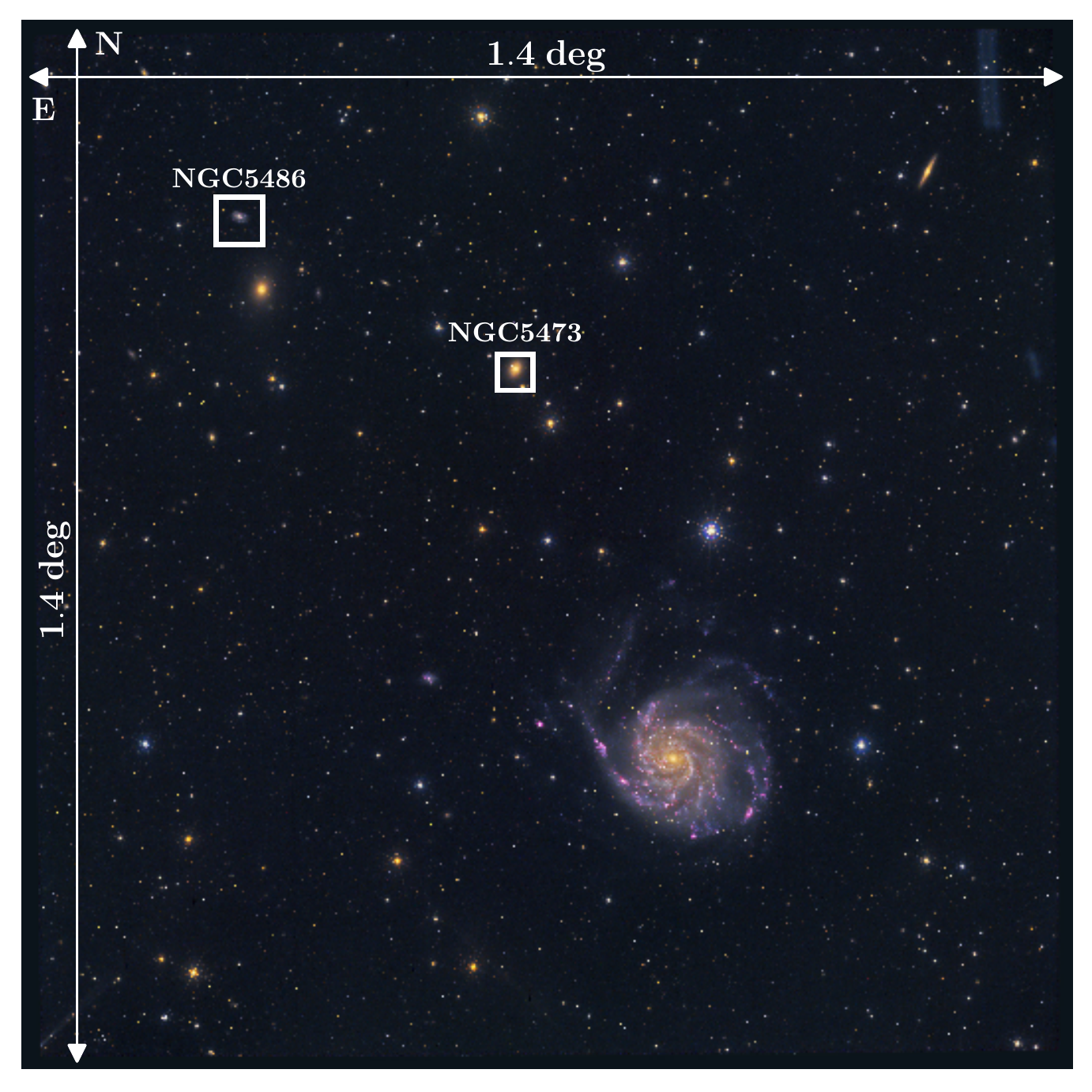}
  \caption{One of the \jplus SVD pointings, including M101 and several NGC galaxies, two of which are indicated in white boxes as they are discussed in this paper.}
  \label{fig:M101}
\end{figure*}

\subsubsection{Photometric redshifts}\label{sec:photoz}

A photo-$z$ is an estimate of the redshift of an extragalactic source based upon its SED as it is inferred from its apparent magnitudes provided by a particular filter system. From its first application in the 1960s \citep{1962IAUS...15..390B}, the estimation of photo-$z$s has experienced a very significant evolution. Nowadays, photo-$z$s have become an essential tool in modern astronomy since they represent a quick and almost inexpensive way of retrieving redshift estimates for a large amount of galaxies in a reasonable amount of observational time. 

Although redshift estimations from galaxy colours are more uncertain than those obtained directly from a spectrum, this situation is certainly being mitigated. The strong dependency between the wavelength resolution (number and type of pass-bands) and the achievable precision of photo-$z$ estimates \citep[e.g.][]{1994MNRAS.267..911H, 2009ApJ...692L...5B} has inspired the design of a whole generation of medium-to-narrow multi-band photometric redshift surveys. Surveys such as COMBO-17, COSMOS, ALHAMBRA or SHARDS have used a combination of broad, medium and narrow-band filters to increase the sensitivity to emission-lines with moderate equivalent widths, providing photo-$z$ estimates as accurate as $\delta_{\rm z}/(1+z)<0.01$ for high signal-to-noise galaxies. As demonstrated in \citet{BenitezJPAS}, the upcoming new generation of $> 50$ narrow-band photometric surveys will push this technique yet further, reaching a photo-$z$ precision level of $\delta_{\rm z}/(1+z) \sim 0.003$ or $0.3$\,\%, equivalent to a resolution of $R\simeq 50$ from a real spectrograph.

\begin{figure*}
  \centering
  \includegraphics[width=0.39\textwidth]{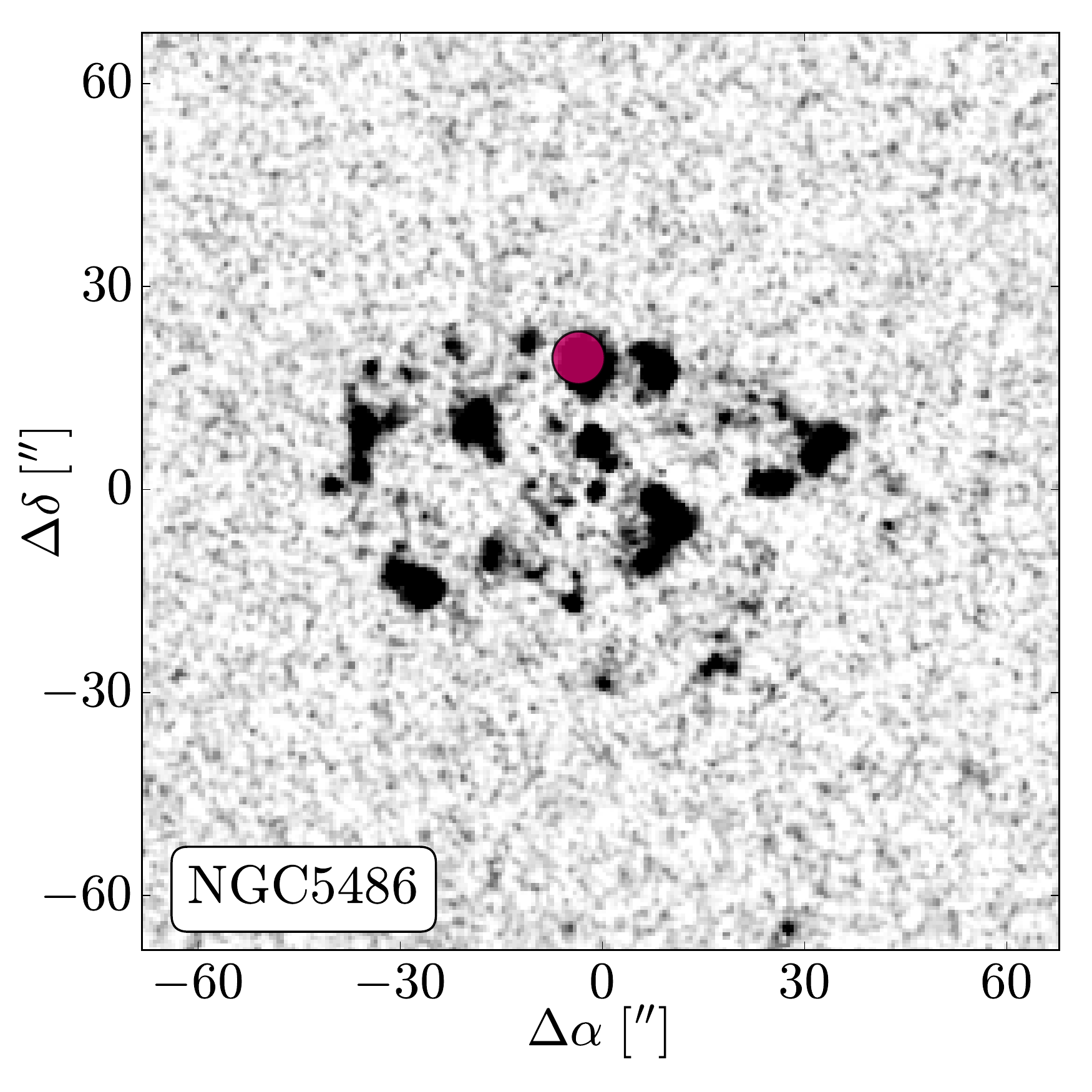}
  \includegraphics[width=0.59\textwidth]{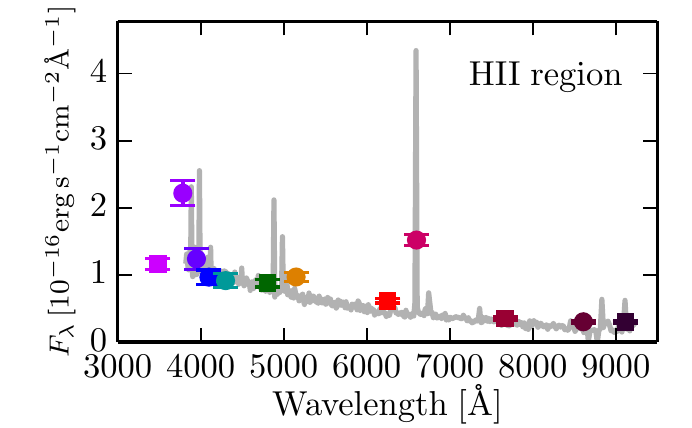}
  \caption{NGC5486 as seen in the \jplus $J0660 - r$ image ({\it left panel}). The panel shows a $53.3$\,arcmin$^2$ area. The red dot marks the position of a HII region observed spectroscopically by SDSS. The \jplus photo-spectrum and the SDSS spectrum of such region are shown in the {\it right panel}. The data sets are normalized to have the same $r-$band magnitude. The $J0660$ and $J0378$ fluxes are raised because of the $H\alpha$ and [\ion{O}{ii}] nebular emission.}
  \label{fig:HII}
\end{figure*}

The main uncertainties in the estimation of photo-$z$s for optical surveys such as \jplus come from i) the available wavelength resolution (number and type of pass-bands), and ii) the limited photometric depth of the observations. The \jplus photometric system is well designed to recover accurate photo-$z$ information of nearby galaxies with $z\lesssim 0.2$, thanks to continuous covering of the 4\,000\,\AA\, break and the presence of particular emission lines. We address the issue of photo-$z$ estimation in \jplus data by comparing photo-$z$ estimates from EDR with spectroscopic redshifts from the literature, in particular from SDSS and from the nearby clusters A2589 \& A2593 (located at $z \sim 0.043$, see Sect.~\ref{sec:2denv} and \citealt{molino17} for further details). We also compare our photo-$z$ estimates with the ``real" redshifts of synthetic, realistically simulated \jplus sources injected in real \jplus images, \citep{molino17}. The codes used in this analysis are {\tt BPZ2}\footnote{\url{http://www.stsci.edu/~dcoe/BPZ/}} and {\tt LePhare}\footnote{\url{http://www.cfht.hawaii.edu/~arnouts/LEPHARE/lephare.html}}.

Let us first introduce a representative example of \jplus photo-spectra for four cluster galaxies located at $z\sim 0.065$ in the EDR tile 4951, see Fig.~\ref{fig:ell_photoz}. These nearby photo-spectra are found to be in good agreement with the outputs from spectroscopy (grey lines). 
We next present a quantitative comparison between \jplus EDR photo-$z$s and SDSS spectroscopic values in Fig.~\ref{fig:photo_vs_spec}. We find that the spectroscopic values are retrieved with no significant bias ($|\Delta z |\equiv |z_{\rm phot} - z_{\rm spec}| < 0.005$ for $r$-band magnitude $<20$), and a typical error $\delta_{\rm z}$ in the range $0.005$--$0.04$ for $r\in [16,20.5]$. These results are in agreement with those from \citet{molino17} in two nearby clusters contained in the \jplus SVD. The explicit dependence of the photo-$z$ bias and precision with the $r$-band magnitude is shown in Fig.~\ref{fig:photozsimul}. As expected, the quality degrades at lower signal-to-noise, reaching $\delta_{\rm z} \sim 0.03$ at $r \sim 20$. We stress that photo-$z$ precision below 1\,\% is reached for galaxies brighter than $r = 18$. We also stress the good agreement with the predictions from the simulations of \citet{molino17} (displayed by squares in Fig.~\ref{fig:photozsimul}). 

Finally, the benefit of extending the classical five broad-band surveys (such as SDSS) to twelve bands including medium width and narrow filters is also shown in Fig.~\ref{fig:photozsimul}. Both the direct comparison with SDSS spectroscopic information and the simulations from \citet{molino17} found a significant improvement in the photo-$z$ precision at all magnitude bins. This reaches a factor as high as 2.5 gain when the signal-to-noise in the narrow-band filters is large enough. Interestingly, as pointed out in \citet{molino17}, the SDSS-like surveys cannot surpass a certain precision in the photo-$z$ estimates irrespectively of the signal-to-noise of the galaxies. This limitation is imposed by the poor wavelength resolution provided by broad-bands, causing a degeneracy in the colour-redshift space. Although \jplus observations are fainter in terms of photometric depth to those of SDSS, they are indeed deeper in terms of photometric redshift depth, due to the addition of the seven narrow-bands to the filter system.

\subsubsection{2D SFR}\label{sec:2dsfr}

As an alternative to spectroscopy, narrow-band filters offer a great opportunity to conduct systematic studies of ELGs, sacrificing the precision that spectroscopy offers for much larger data samples with significantly better statistics. The SFR is known to correlate with the emission of H$\alpha$ by the heated gas in star-forming regions \citep[][]{Kennicutt1998}. Thanks to the large sky coverage and the specific rest-frame H$\alpha$ filter $J0660$, \jplus is a very powerful survey to study the integrated and the spatially resolved SFR at $z<0.015$.

The work of \cite{vilella15} describes an optimized methodology to extract the H$\alpha$ flux and the SFR from \jplus-like photometric data. First, the H$\alpha$+[\ion{N}{ii}] flux covered by the $J0660$ filter is isolated by substracting an underlying continuum estimate. This estimation is produced by a SED-fitting technique that uses all the available \jplus filter set. After that, the dust extinction is corrected for and the [\ion{N}{ii}] flux removed using empirical relations derived from the SDSS spectroscopic sample. We test the above procedure with synthetic photometry derived from the SDSS spectra of star-forming regions, and find that the measured H$\alpha$ flux is unbiased. This is in contrast with usual methods based solely on $J0660 - r$ information, which underestimate measurements of H$\alpha$ fluxes by $\sim 20$\% \citep{vilella15}.

We also check this procedure based on synthetic photometry with \jplus SVD and EDR, as presented in \citet{logronho17}. We compare the H$\alpha$ flux measured with \jplus photometry against spectroscopic measurements from SDSS and CALIFA in 46 shared HII regions. We find that the \jplus H$\alpha$ flux is consistent with the spectroscopic one, with a median flux ratio $R = F_{{\rm H}\alpha}^{\rm J-PLUS} / F_{{\rm H}\alpha}^{\rm spec} = 1.01 \pm 0.04$ \citep{logronho17}. This demonstrates the capabilities of the \jplus filter set and validates the methodology presented in \cite{vilella15}.

The M101 \jplus SVD field is presented in Fig.~\ref{fig:M101}. In addition to M101, other nearby galaxies are presented in the field, such as the elliptical galaxy NGC5473 (see Sect.~\ref{sec:2dsps}) and the star-forming galaxy NGC5486. To illustrate \jplus potential on the study of H$\alpha$ emission, we present the study of one NGC5486 HII region in Fig.~\ref{fig:HII}. \jplus photometry at the location of the SDSS fibre is consistent with the SDSS spectra, and presents a clear flux excess in the $J0660$ band due to the H$\alpha$+[\ion{N}{ii}] emission of the region. The ratio between the \jplus and the spectroscopic H$\alpha$ flux is $R = 0.86 \pm 0.18$, i.e., compatible with unity. Moreover, the $J0378$ band is also raised because of the nebular [\ion{O}{ii}] emission. This line is outside the SDSS spectral coverage at $z < 0.015$, while \jplus provides valuable information for systematic [\ion{O}{ii}] studies in the nearby Universe. Finally, the \jplus $J0660 - r$ image of NGC5486 shows a population of HII regions lacking spectral information. The study of spatial gradients within those galaxies, together with individual HII regions in a large, nearby galaxy sample will be possible after \jplus completion. As a bonus, it will also be possible to address the impact of the environment on those systems.

\subsubsection{2D stellar populations}\label{sec:2dsps}

The spatial variations of stellar population properties within a galaxy are intimately linked to their formation process.  Therefore, spatially resolved studies of galaxies are critical to uncover the history of formation and assembly of local galaxies. Although the arrival of integral field spectroscopy (IFS) surveys constitutes a significant breakthrough in the field,  recent techniques that combine photometric multi-filter surveys  with spectral fitting diagnostics have opened a new  way  to  disentangle   the   stellar   population   of   unresolved   extended galaxies, and with a relatively low-cost compared to IFU surveys. 

Current generations of IFS surveys allow detailed internal analyses through multiple spectra of each galaxy by creating a 2D map of the object. While these surveys are very powerful, there are still limitations in terms of accesible redshift ranges and galactocentric distances. For example, MaNGA aims to obtain spatially resolved spectroscopy of 10\,000 galaxies but it will be limited to resolve galaxies spatially out to $R=1.5\,$R$_\mathrm{eff}$\footnote{The effective radius R$_{\mathrm{eff}}$ is defined as the radius containing half of the total light emitted by the galaxy.}  (with a subsample reaching $R=2.5\,$R$_\mathrm{eff}$), and with a median redshift of $z\sim 0.03$. For the SAMI survey, redshifts are limited to $z < 0.095$ and the data typically reaches $1.7\,$R$_\mathrm{eff}$ ($2\,$R$_\mathrm{eff}$ for 40\,\% of the sample).

\citet{Sanromanetal2018a} developed a technique to analyze unresolved stellar populations of spatially resolved galaxies based on photometric multi-filter surveys. They derived spatially resolved stellar population properties and radial gradients by applying a Centroidal Voronoi Tesselation and performing a multi-color photometry SED fitting \citep{MUFFIT}.  The method has been tested and validated on a sample of 29 early-type galaxies observed by ALHAMBRA survey using 23 medium filter bands at the 3.5\,m telescope of Calar Alto observatory. This observing technique enables spatially resolved stellar population studies out to considerably fainter surface brightness levels that are not possible with current IFU surveys. It also allows studies of very nearby galaxies ($z < 0.01$) that are so spatially extended to be unsuitable for the small FoV of current IFU surveys. In this context multi-filter surveys open a way to improve our knowledge of galaxy formation and evolution that complements standard multi-object spectroscopic surveys. 

\begin{figure*}[h]
\begin{center}

\includegraphics[width=1.0\textwidth]{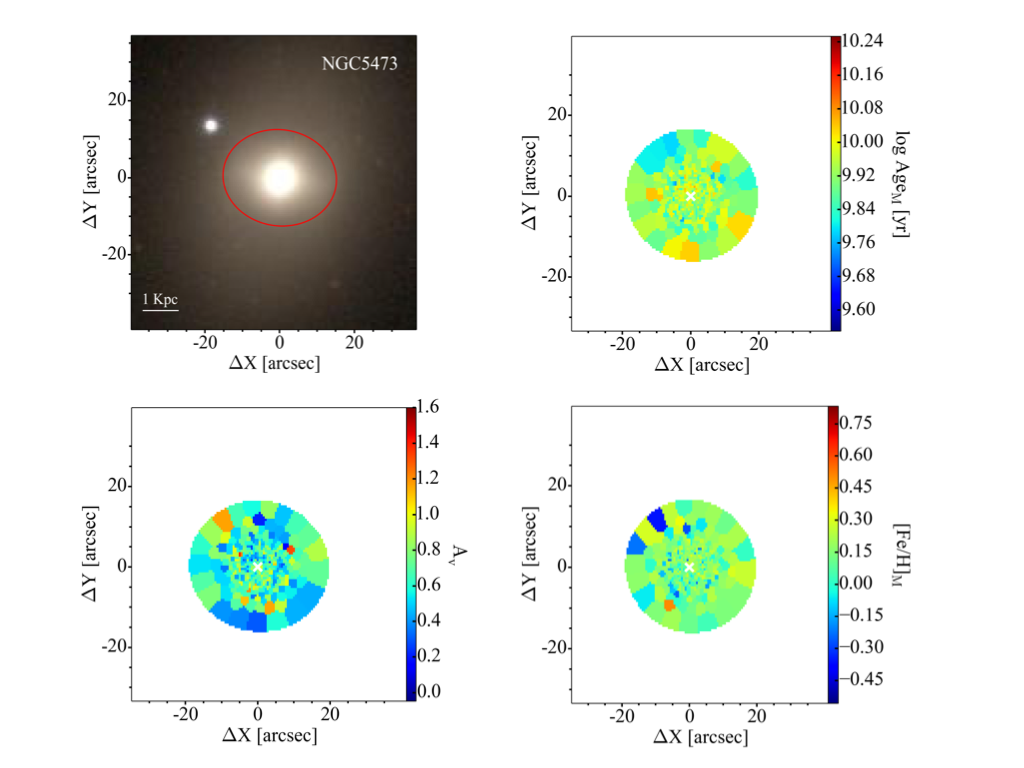}

\caption{Illustrative example of the 2D methodology. The top left panel shows a \jplus colored composite image of the galaxy NGC5473. The red ellipse delimits a ellipse with the semi-major axis $R=3\,$R$_\mathrm{eff}$. The rest of the panels show the mass-weighted stellar population properties maps ($\log\mathrm{Age}$, A$_\mathrm{v}$ and [Fe/H]) for this object.}
\label{fig:2Dstudy}
\end{center}
\end{figure*}

In order to explore the IFU-like potential of \jplus, we have applied the previous multi-filter technique to two galaxies in common with CALIFA. As an illustrative example, Fig.~\ref{fig:2Dstudy} presents the age, metallicity and extinction maps of the elliptical galaxy NGC5473. Detail inspection of the 2D maps and the radial profiles show a flat age gradient and a mild negative metallicity gradient (i.e., the inner part is more metal-rich than the outer part of the galaxy). These results are in agreement with previous long-slit analysis  \citep[e.g.,][]{SanchezBlazquezetal2006, SanchezBlazquezetal2007, Spolaoretal2010}, and also with the most recent IFU studies \citep[e.g.,][]{Kuntschneretal2010, Wilkinsonetal2015, Goddardetal2016}. A detailed analysis, including a one-to-one comparison with different IFU techniques, is presented in a companion paper \citep{sanroman18}. 

\subsubsection{Environmental studies in the nearby Universe}\label{sec:2denv}
One of the main strengths of large-area photometric surveys is the continuous information provided across the field. This enables the analysis of environmental effects, ranging from close pairs and nearby interactions to clusters and large scale structure.


The photometric redshift accuracy reached by \jplus in the nearby
universe, with an uncertainty level $\delta_{z} \sim 0.02$
(Sect.~\ref{sec:photoz}), will improve our knowledge about local
groups and clusters, and about their galaxy members. The potential of
\jplus in this topic is demonstrated in \citet{molino17}, where \jplus
SVD of the two nearby galaxy clusters A2589
($z=0.0414$) and A2593 ($z=0.044$) are analysed. Three JAST/T80Cam
pointings were required to observe both clusters and the region
between them, since they are separated by 3.5\,deg on the sky. The
photometric redshifts measured with \texttt{BPZ2} are consistent with
the EDR estimations presented in Sect.~\ref{sec:photoz}, as well as
with the \jplus forecast based on the simulations performed
by \citet{molino17}. The empirical redshift probability distribution
functions (PDFs) can be used to look for potential new faint members
in these clusters, providing a membership probability for each
observed source
\citep[see][and its Fig.~9 therein for further details]{molino17}. This approach is
reassured by the good agreement found between our PDFs and the
spectroscopic redshifs for those galaxies for which spectra are
available in the literature. This indiscriminate, high quality, photo-$z$ estimation for galaxies in the cluster fields opens the way to statistical studies of
cluster membership, while providing valuable data for optimal target
selection in spectroscopic follow-ups. In particular, photo-$z$s
produced by \jplus are going to be used by the WEAVE cluster surveys
in the process of target selection. It is expected that \jplus photo-$z$s will increase the number of selected cluster galaxy members by a factor $2$--$3$. This improvement is particularly relevant in the range of faint magnitudes and in the outskirts of galaxy clusters.

\begin{figure}
\centering
\includegraphics[width=9cm]{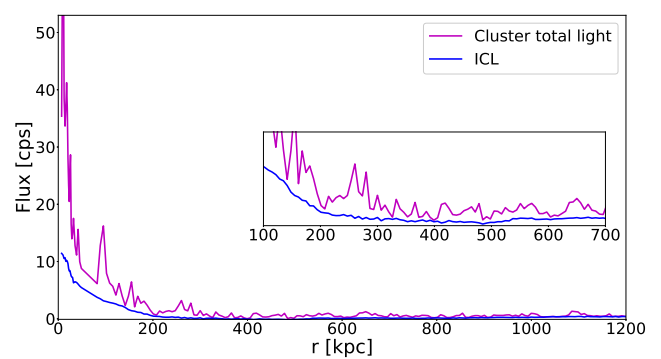}
\caption{Radial profiles of the cluster total light (red), and the ICL (blue) in the $J0410$ band.}
\label{fig3ICL}
\end{figure}

\jplus is also suitable to study the intracluster light (ICL), generally defined as the luminous component of galaxy clusters belonging to stars that are gravitationally bound to the cluster gravitational potential, but not locked in any of the individual member galaxies. They are believed to be released through dynamical stripping of stars during the hierarchical process of cluster assembly and growth \citep{2014MNRAS.437.3787C}. The ICL fraction and its properties are thus key to understand the formation history of galaxy clusters \citep[e.g.][]{2004ApJ...609..617F,2011ApJ...732...48R,2014ApJ...794..137M}, as well as to determine the correct baryon fraction to be compared to cosmological estimates \citep[e.g.,][]{2004ApJ...617..879L}.

Methods employed in the literature are based on binding energy, local density or surface brightness cut offs. These techniques yield different results in the ICL fraction, with discrepancies up to a factor of four using the same data \citep{2011ApJ...732...48R}. For this reason, we have developed a method to measure ICL fraction (defined as the ratio between the ICL flux and the total light of the cluster), which is free from {\it a priori} assumptions of surface brightness profiles and cutoffs \citep{2016ApJ...820...49J}. The method is called {\tt CICLE} (CHEFs ICL Estimator), where the Chebyshev-Fourier functions (CHEFs, \citealt{2012ApJ...745..150J}) are a set of very efficient and compact orthonormal mathematical bases that are able to model galaxies of any morphological type. The CHEFs are used in {\tt CICLE} to model and remove the light from the galaxies in the field, including the brightest cluster galaxy (BCG). The special configuration of the \jplus filter system and the large contiguous area covered by the survey makes it possible to study for the first time the ICL colors of nearby clusters using narrow bands. This yields low resolution information on the spectral energy distribution of the ICL, and provides key information about the origin of this stellar component. As a first example, we show the preliminary results of {\tt CICLE} implemented on \jplus $J0410$ observations of the Coma cluster, obtaining the profiles shown in Fig.~\ref{fig3ICL} for the ICL and the total luminosity of the cluster. We find a final ICL fraction of 24\,\% for a radius of $\approx 490$\,kpc. This is in agreement with previous works and simulations on Coma \citep[][]{melnick1977,bernstein1995,adami2005}. Further details are to be found in Jim\'enez-Teja et al. (in preparation).

\subsection{Redshift windows in the distant Universe}
\label{sec:zwin}

The filter set configuration of \jplus, including several narrow band filters, allows for a unique search of high redshift targets displaying emission lines. The wide area coverage of the survey compensates its shallow magnitude limit, thus opening a window to explore the abundance and properties of statistical samples of bright ELGs at different redshifts. Many redshift windows provided by the narrow band filters can be used to target a strong optical emission line. Most notably, i) $z \sim 0.77$, when the [\ion{O}{ii}] $\lambda \lambda 3726,3729$ doublet enters the $J0660$ filter; ii) $z \approx 2.1\,(2.2) $, when the Ly$\alpha$ line enters the $J0378\,(J0395)$ filter, and iii) $z\sim 4.4$, when Ly$\alpha$ line enters the $J0660$ filter. 

To identify an emitter and measure its line flux it is necessary to detect a magnitude excess in the narrow-band filter with respect to the adjacent filters that trace the continuum around the line. In the \jplus filter configuration, the continuum around a line can be estimated using a combination of the \jplus broad-band filters. \citet{vilella15} developed a simple, yet robust methodology to extract the H$\alpha$ flux from local star-forming galaxies using either a combination of three \jplus filters (two broad bands and one narrow band), or all the filters via SED fitting (see Sect.~\ref{sec:2dsfr} and \citealt{logronho17}). These methods can be adapted to other lines at other redshifts.

Typical searches for ELGs using narrow bands are significantly affected by interlopers, i.e., line confusion. With \jplus data, all the twelve filters can be used to gain color information that mitigate the impact of emitters at undesired redshifts.

\begin{figure}
\centering
\includegraphics[width=0.5\textwidth]{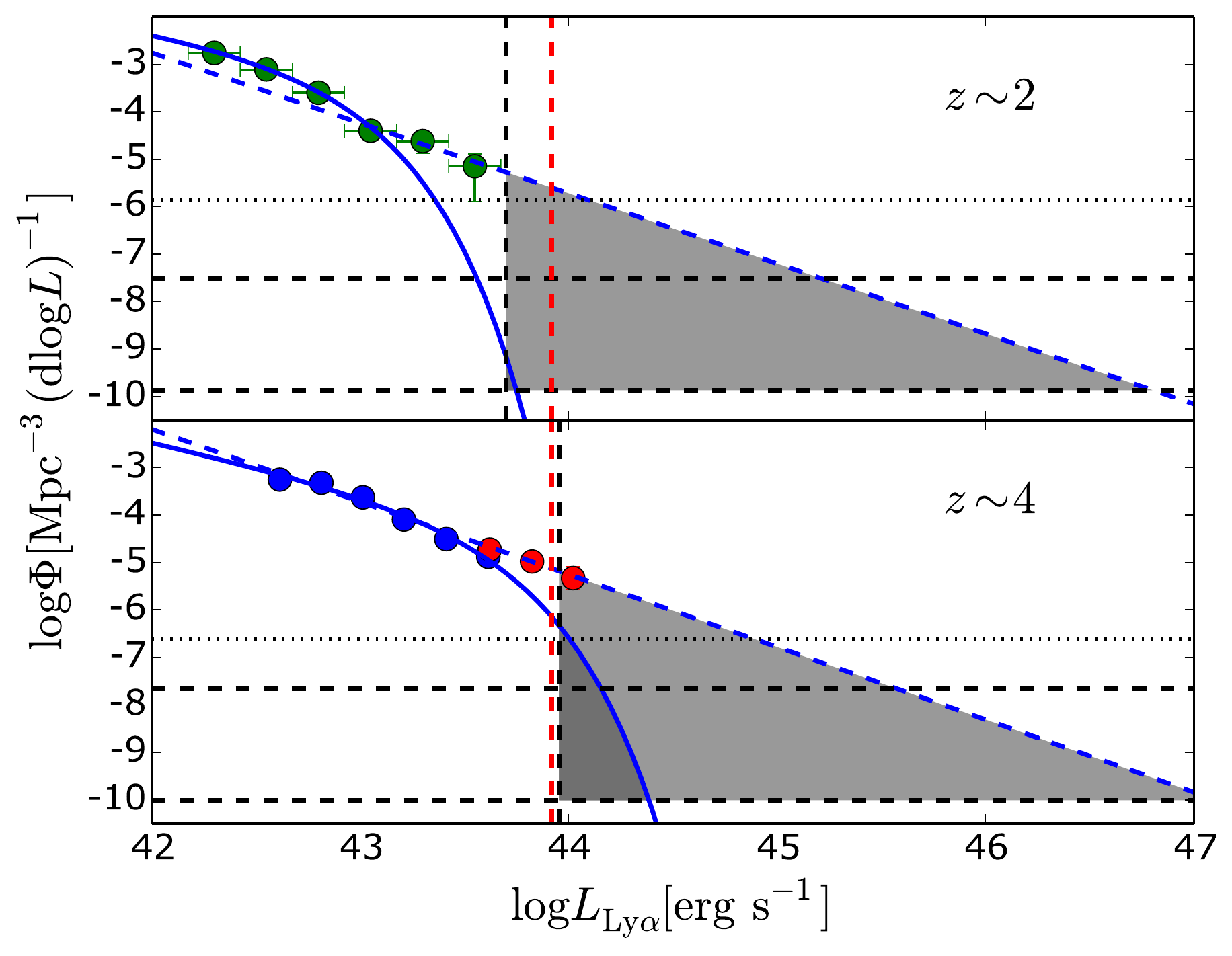}
\caption{
The bright-end of the LAE luminosity function (LF) accessible by \jplus. The gray shaded areas in both panels show the region of the LF that can be probed with the \jplus area and flux depth (dark grey for a Schechter-like LF, light gray for a power-law LF). The horizontal dotted/dashed lines represent the lowest possible value of the LF achievable with the \jplus EDR/full survey. The black vertical dashed line shows the limiting luminosity of \jplus at redshifts 2.2 ({\it top}) and 4.4 ({\it bottom}). The vertical red dashed line shows the luminosity of CR7 at $z= 6.604$ \citep{sobral15} for illustration. Green circles in the top panel show the Ly$\alpha$ LF from \citet{sobral17}. Blue and red circles correspond to the UDS+COSMOS and SA22 samples from \citet{matthee15}. The solid blue and dashed lines correspond to Schechter and power-law fits to the data, respectively.
%
%
}
\label{fig:be2_jplus}
\end{figure}

\subsubsection{[OII] emission at z $= 0.77$}
ELGs at $z \sim 0.77$ are currently being targeted by several ongoing and planned Multi-object spectroscopic (MOS) 
cosmological surveys, such as 4MOST, DESI, PFS and eBOSS \citep{dejong12,weinberg13,takada14,delubac17}. Broad-band target selection designed to select ELGs at $z\sim 0.77$ can deliver success rates of the order of 70\,\% due to the completeness and purity of the broad band photometry used to define their targets \citep{raichoor17}. Since \jplus is a photometric survey, its selection function depends only on the depth of the images and the ability to assign the correct redshift to an object with flux excess in a narrow band filter. Thus, the characterization of the ELG population with \jplus data is expected to be highly complementary to that coming from MOS facilities.

Assuming a minimum observed equivalent width EW$ = 100$\,\AA, we estimate that we can reach the [\ion{O}{ii}] line emission down to the luminosity of $\sim 1 \times 10^{42}$\,erg\,s$^{-1}$. These objects will be used to characterize the abundance and clustering properties of star-forming galaxies at $z=0.77$ (Orsi et al., in prep.)

\subsubsection{Ly$\alpha$ emission at z $= 2.2$ and $4.4$}

At higher redshifts, detections are strongly limited by the survey
depth. Star-forming Ly$\alpha$ emitting
galaxies (Ly$\alpha$ emitters, or LAEs in short) at $z\sim2$ are
expected to be detected down to Ly$\alpha$ luminosities of
$\sim 5\times 10^{43}$\,erg\,s$^{-1}$ assuming a minimum observed EW of $100$\,\AA. At $z = 4.4$, we expect to detect LAEs down to
$\sim 9\times 10^{43}$\,erg\,s$^{-1}$.

\citet{sobral15} have confirmed spectroscopically an extremely
luminous LAE at $z=6.604$, named CR7, with Ly$\alpha$ luminosity of
$8.3\times 10^{43}$\,erg\,s$^{-1}$. A thorough census of these extreme objects
has not been yet possible due to the prohibitively large volume required. 
The \jplus dataset can thus perform an unprecedented search for these
extremely bright emitters at high redshifts (Fig.~\ref{fig:be2_jplus}). Furthermore, \jplus is also expected to detect Ly$\alpha$ blobs (LABs)
among the usual high-$z$, star-forming, Ly$\alpha$ emitting, compact
galaxies. LABs are an extremely rare class of Ly$\alpha$ emitters
(comoving number density $n_c<10^{-6}$\,Mpc$^{-4}$) showing spatially
extended Ly$\alpha$ emission ($\sim30-200$\,kpc in size, see
e.g.,s \citealt{smaila2016}).

\begin{figure}
\centering
\includegraphics[width=0.49\textwidth]{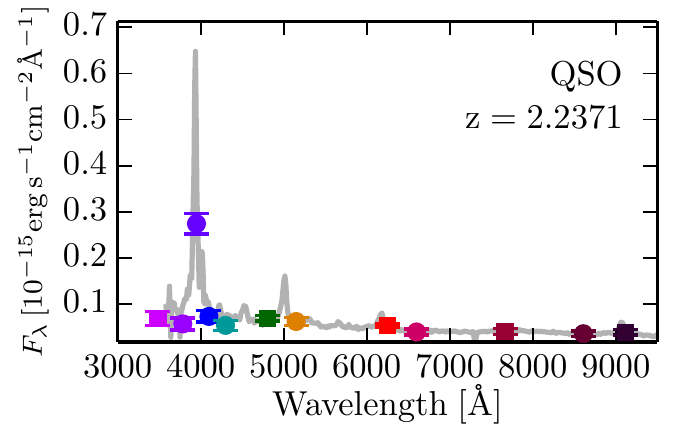}\\
\includegraphics[width=0.49\textwidth]{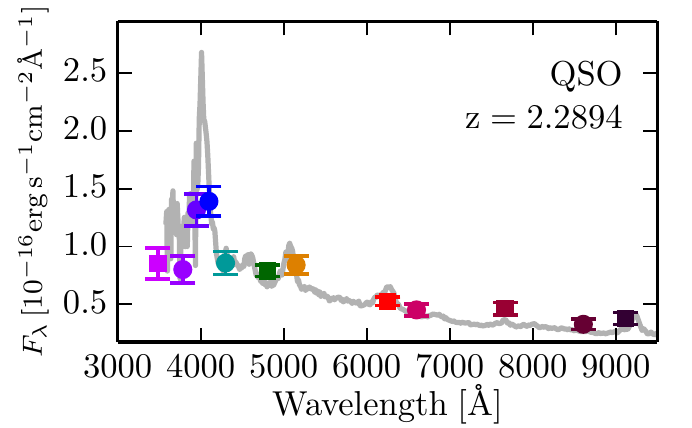}
\caption{\jplus photo-spectrum of two QSOs with SDSS spectra (marked in Fig.~\ref{fig:fig_edr}),
located at $\alpha = 110.11167$, $\delta = 40.23105$, $z= 2.2371$ ({\it top panel}), and 
$\alpha = 110.22740$ , $\delta = 40.14428$, $z= 2.2894$ ({\it bottom panel}).
The Ly$\alpha$ broad-line emission is clear in both sources and is captured by the $J0410$ filter, 
showing a relative excess with respect to the continuum traced by both broad-bands and rightmost 
narrow-band filters.}
\label{fig:jplus_qso}
\end{figure}

One straightforward approach to select LAE candidates is the three-filter method of \citet{vilella15}
over the filter sets [$u$, $J0378$, $g$] and [$u$, $J0395$, $r$]. This method, 
combined with a selection criteria to assess the
narrow-band excess significance (see, e.g., \citealt{guaita2010, konno2016}), 
allows to easily extract LAE candidates within \jplus EDR (Spinoso et al., in
prep.). However, the emission in a single \jplus narrow-band is not
sufficient to distinguish among the different sources of Ly$\alpha$ emission
(star forming galaxies and quasars), as well as to reject interlopers at other redshifts.  
Quasars constitute a large fraction of the LAE population detectable at
the depth of \jplus (Fig.~\ref{fig:jplus_qso}). While a significant number of $z\sim 2$ quasars has already been observed within SDSS spectroscopic programs, we expect to be able to
identify candidate quasars missed by automatized spectroscopic target selection
provided that we simply select candidates based on their
Ly$\alpha$ emission. This may turn out to be particularly relevant for quasars in the range of magnitudes $r=12$--$15$, where SDSS stellar images are usually saturated and \jplus photometry can play a unique role. 

By exploiting the complete \jplus filter set, interlopers
at undesired redshifts can be indentified by the presence of additional
narrow-band excesses other than the one in $J0378$ or $J0395$, and removed. This
method relies on the assumption that compact star-forming Ly$\alpha$ emitting
galaxies at $z\sim2$ should only exhibit strong line emission at Ly$\alpha$ wavelengths. Our preliminary study also indicates that some combinations of \jplus broad-band colours are useful in removing the low-$z$ contaminants.
Finally, an additional way to reduce the contamination
from line emitters at $z<2$  is to cross-match the \jplus
candidates  with other public available catalogues. Assuming that 
high-$z$ LAEs should not exhibit strong
emission below the Lyman break, all the sources with a counterpart in the
GALEX catalogue can be excluded from our selection. Other undesired
contaminants can be removed by cross-matching \jplus sources with the SDSS spectroscopically-identified samples of stars and low-$z$ galaxies.

In summary, the \jplus filter configuration allows to select samples of
emission-line galaxies and quasars in narrow redshift ranges when the line
emission can be isolated and characterized. From a preliminary analysis of the
\jplus EDR, for example, we selected a sample of  $\sim30$ LAE candidates consistent with $z\approx2$ Ly$\alpha$ emitting sources, including galaxies and QSOs, which we
plan to carefully analyze with spectroscopic follow-up.

\subsection{Variable sources}\label{sec:var}
\begin{figure*}
  \centering
  \includegraphics[width=\textwidth]{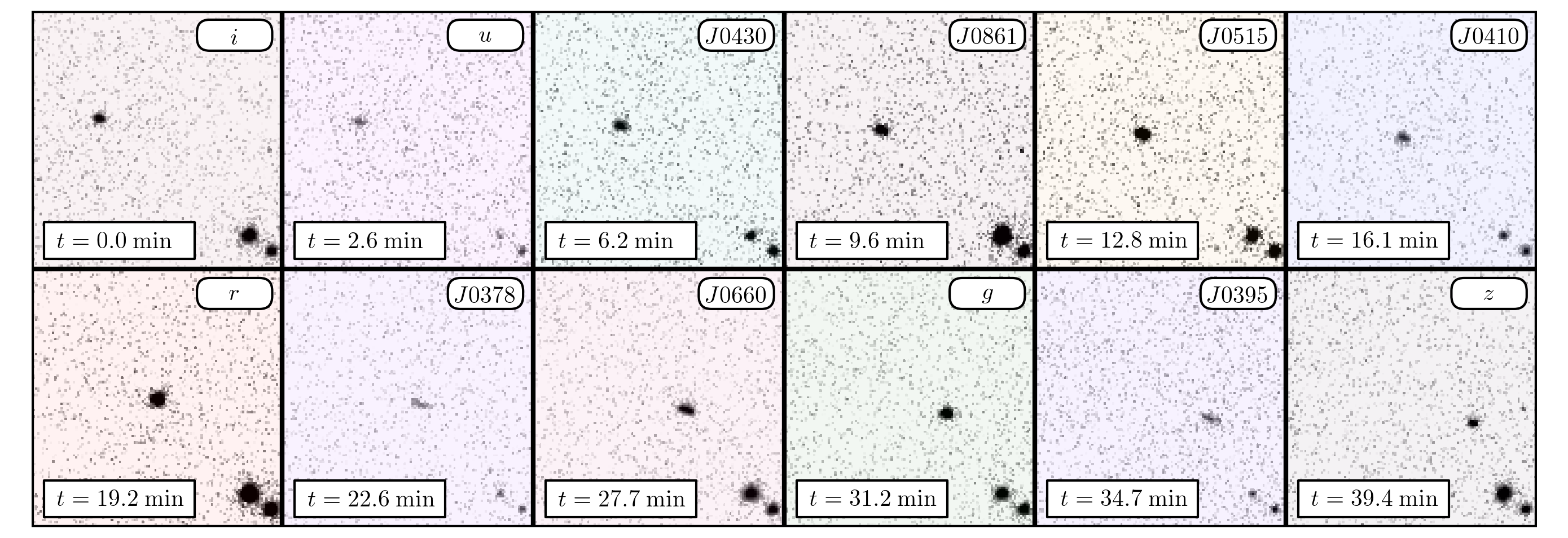}
  \caption{The 12 \jplus band images of the minor body marked in Fig.~\ref{fig:fig_edr}. The observing sequence spans 39.4\,min, starting with the $i$ band and finishing with the $z$ band.}
  \label{fig:MB_filters}
\end{figure*}

Exposures on each \jplus filter are done sequentially, and this provides sensitivity to source variability (either in flux and/or sky position) on time scales shorter than $\sim 1$\,hour.  This time scale corresponds to the typical time interval between the first and the last exposure of the same pointing, taking into account all the filters. A clear example for the position variability of a minor body in our solar system can be found in Fig.~\ref{fig:MB_filters}. \jplus will hence trace objects that vary their position, such as Solar System minor bodies (Sect.~\ref{sec:mb}), and their flux, such as cataclysmic variables (Sect.~\ref{sec:cvs}) and variable RR Lyrae stars (Sect.~\ref{sec:rrl}). 

\subsubsection{Minor bodies of the Solar System}\label{sec:mb}

All sky photometric surveys like \jplus will necessarily observe a
large number of minor bodies of the Solar System  during normal operation. 
The majority of the known minor bodies are found in the Main Asteroid
Belt, with their orbits confined between Mars' and Jupiter's. Their
compositions have a direct relation to the region of the
protoplanetary disk where they formed. Bodies that were formed beyond
the snow line - the distance from the proto-sun beyond which water ice
is stable - would have accreted a considerable amount of ice-rich
dust.

The presence of hydrated material in the Main Asteroid Belt has been
inferred through the analysis of their reflectance spectra and colours
thanks to the presence in the visible spectra of the asteroids of a
broad, shallow phyllosilicate band centered around $0.7\,\mu$m. The
presence of this band indicates not only that the material were
originally rich in water ice, but also that the interior of the body
reached temperatures capable of sustaining the presence of liquid
water for an extended period of time.  The occurrence of this
  band in asteroids has been mapped using low dispersion
  spectroscopy, showing that the fraction of asteroids where these
bands are observed varies with heliocentric distance, with lower
limits of 10\,\% of the observed asteroids at 2.3\,AU. This percentage
increases to 25\,\% at 3.1\,AU and falls towards the outer edge of the
Main Asteroid Belt \citep{2003Icar..161..356C}.

\begin{center}
\begin{figure}
\includegraphics[width=0.5\textwidth]{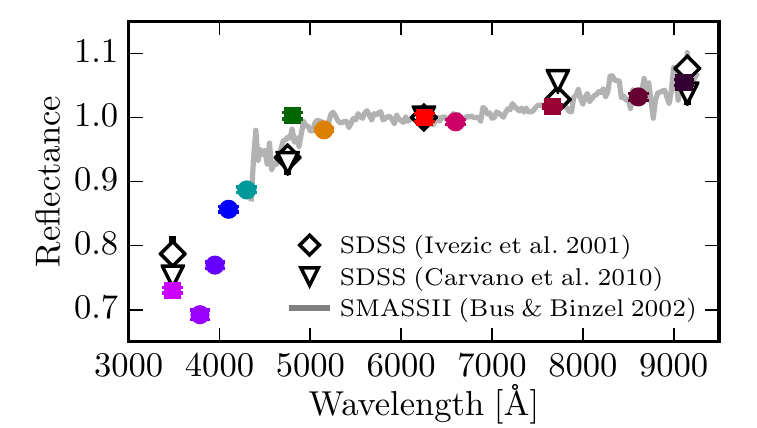}
\caption{Reflectance spectrum of the hydrated asteroid (1024)
  Hale obtained from \jplus observations, together with two sets
  of reflectance spectra from SDSS observations
  \citep{2001AJ....122.2749I,2010A&A...510A..43C}  and other obtained
  with low dispersion spectroscopy \citep{2002Icar..158..106B}. All
  spectra normalized at the $r$ band. The \jplus observations were
  made during normal survey operation on 2017-01-07 while the asteroid had
  $V=15.2$\,mag. The solar colours used to obtain the \jplus reflectance
  spectrum were calculated using a reference solar spectrum from
  \cite{2010JQSRT.111.1289C}. The resulting error bars are smaller
  than the size of the points in the figure. }
\label{fig:conv}
\end{figure}
\end{center}

However, all these data are
based on the relatively small sample of asteroids with spectroscopic
observations. Although most asteroid taxonomical classes can be defined using
intermediate band filters like the  SDSS filter system
\citep{2010A&A...510A..43C},  the low spectral resolution  tends to
render the detection of the $0.7\,\mu$m band unreliable using SDSS data
\citep{2012Icar..221..744R}. In this regard, the larger spectral
resolution provided by the \jplus filter system can provide a better
definition of this band, since with \jplus the spectral interval
covered by the band is sampled also by 2 narrow band filters in
addition to the $r$, $i$, and $z$ filters. This
can be seen on the reflectance spectrum of the asteroid (1024) Hale,
obtained from observations taken during normal survey conditions, see Fig.~\ref{fig:conv}.
 
\begin{center}
\begin{figure}
\includegraphics[width=0.5\textwidth]{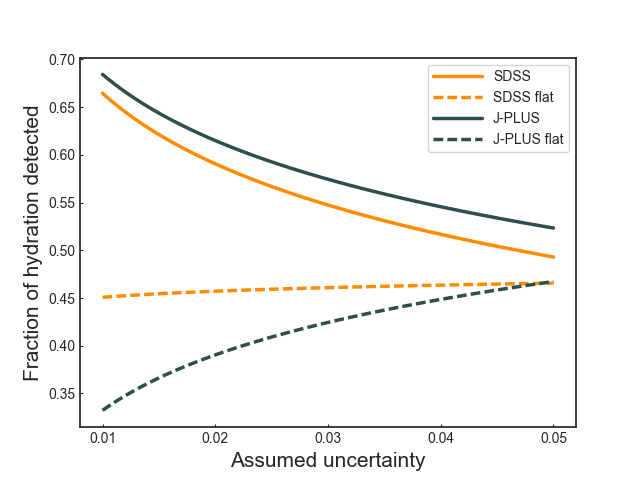}
\caption{Fraction of clones where the algorithm 
  reported the presence of a band using SDSS and \jplus filters as a
  function of the assumed uncertainty of the reflectances. The
  curves corresponding to hydrated template spectra are drawn with
  solid lines, while those corresponding to flat spectra templates are given by 
  the dashed lines.}
\label{fig:frac}
\end{figure}
\end{center}

One way to compare and quantify the sensitivity of the SDSS and \jplus
filter systems to the $0.7\,\mu$m band is to use low resolution
spectra of asteroids with this feature as templates to produce
synthetic reflectances for each filter. We use 5 spectra of asteroids
from the hydrated asteroid family of Erigone to produce a sample of
1000 synthetic spectra with Gaussian random uncertainties in
reflectance driven around the nominal reflectance values and with
nominal RMS. A similar set of clones were generated using a synthetic
flat spectra to simulate a non-hydrated asteroid. To define whether
each clone had a band or not, we generate further 1000 instances of
the measurements by again randomly sampling reflectances normally
distributed around each reflectance measured value. We next
measure, for each realization, the parameter associated to the
presence of the band. The distribution of the parameters are then
compared using a Kolmogorov-Smirnov (K-S) test to the distribution
obtained from featureless reference spectra with the same
uncertainties. We accept the band as real if the K-S test indicates
that the distributions are statistically different and if the
parameter associated with the band presence is larger than a given
threshold. This exercise is performed using the SDSS and \jplus filter
systems. For the SDSS, the parameter associated with the presence of the band is the
reflectance at $i$ on the spectra normalized by the gradient of
reflectance between $g$ and $z$ (with values above that
continua set as negative). For \jplus this parameter is the sum of
area of the polygon formed by the segment that connects the
$r$, $J0660$, $i$, $J0861$, and $z$ bands, and the continuum
defined by the segment between the $r$ and $z$ filters,
with areas above that line counted as negative. Figure~\ref{fig:frac}
shows the fraction of clones in which band detection occurred among
the samples corresponding to templates with (solid lines) and without
the feature (dashed lines), using the SDSS and the \jplus filter sets,
as a function of the assumed uncertainty of the reflectances. This
result shows that the \jplus filter set is more effective in detecting
the band than the SDSS filter set. In particular, the results obtained
with the \jplus filter set present a considerable smaller number of
false positive in comparison with data obtained with the SDSS filters,
with the difference as high as $\sim 29\,\%$ less false detections for $1\,\%$ uncertainties in the reflectances.

Using data from the observations obtained so far it is possible to
estimate the depth at which asteroids can be  expected to be observed
during the survey. 
We find that the detection
efficiency at the bluer filters starts to fall around $V=17.5$\,mag, but for
the filters involved in the detection of the $0.7\,\mu$m band the
detection efficiency remains high up to $V=19$\,mag. Simulations
indicate that during the survey execution 
time over ten thousands known asteroids brighter than $V=19$\,mag will be
observed. Therefore, once completed the \jplus data set of minor
bodies will provide a valuable tool to probe the distribution of
hydration in the Main Asteroid Belt.

\subsubsection{Searching for Cataclysmic Variables}\label{sec:cvs}

\begin{figure}
	\centering
	\includegraphics[width=0.5\textwidth]{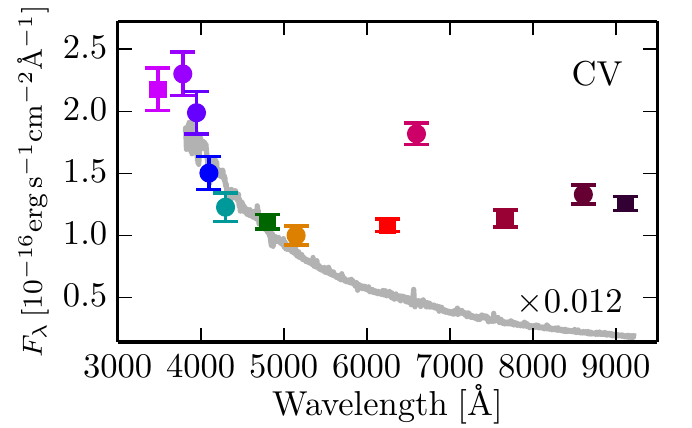}
\caption{\label{fig:cvs}
\jplus photo-spectrum of the cataclysmic variable J0743+4106. The blue side of the spectrum
is dominated by the white dwarf emission while the red side is dominated by the secondary star. 
The $J0660$ filter shows the prominent
$H\alpha$ emission of the disc. For comparison, the spectrum from SDSS
DR12 is shown in gray (but multiplied by a factor 0.012 to match the bluest filters).
This spectrum was clearly obtained during a high mass-accretion phase
while the \jplus photometry was obtained during a low mass-accretion 
phase.}
\end{figure}

Cataclysmic variables (CVs) are interacting binary stars made of a white dwarf and a main sequence star which is loosing mass through Roche lobe overflow \citep[for a review, see][]{warner1995}. 
Historically, CVs have been discovered through their variability properties and, most recently, as by-products of quasar searches \citep[for a review, see][]{gaensicke2005}. In fact, SDSS discovered more than 400 new CVs \citep{szkody2011}.
Depending on the mass transfer rate, the magnetic activity of the white dwarf and the inclination of the system CVs can display a variety of observational properties. Therefore, in general terms, a CV spectrum can be blue and/or red and due to the accretion disk it uses to show $H_\alpha$ emission. 
We checked that from all the CVs of the catalogue by \citet{ritterkolb2003} only one (J0743+4106) falls within the EDR footprint (shown in Fig.~\ref{fig:cvs}). The SED clearly shows the blue and red appearance due to the two stars of the binary as well as the $H_\alpha$ emission of the accretion disc. Interestingly, the SDSS spectrum looks completely different, being clearly dominated
by the continuum of the disc (as would be expected in a high
mass-accretion phase). In Abril et al. (in prep.), we propose a new methodology, based on the \jplus filter set, which can be used to separate CVs from other type of astrophysical sources,
in particular quasars. Our expectations are that \jplus will detect around 4\,000 CVs.

\subsubsection{RR Lyrae stars}\label{sec:rrl}
RR Lyrae stars are evolved low mass stars burning He in their core that
can be found in a region of the Hertzsprung–Russel Diagram where the
Horizontal Branch (HB) crosses the instability strip. These type
of stars undergo radial oscillations and they are classified according
to the excited oscillation mode they suffer: fundamental mode
oscillators only (RRab type), first overtone oscillators (RRc) or
fundamental and first overtone oscillators (RRd).

The large and indiscriminate area \jplus will observe,
together with the survey's depth, makes it very convenient for
deriving properties of the Galactic halo structure. Among the stars than
can be used for that purpose, RR Lyrae stars are of outstanding importance
for several reasons: i) they are ubiquitous
species in our Galaxy, so they
can be found distributed virtually everywhere without being
linked to any particular Galactic component; ii) they are relatively bright
($M_{\rm{V}}\simeq0.6$ for mean halo metallicity), so they are easily
detectable up to a few hundred kpc from us; iii) their pulsation periods
obey a period-luminosity-metallicity relation that makes them standard
candles, becoming very useful to constraint distances; iv) they are
relatively old stars, so they are fair tracers of the MW old component.
For all these reasons RR Lyrae stars are excellent tracers of the
structure of the Galactic halo (e.g. \citealt{Sarajedini2011}).

\begin{figure}[t]
	\centering \resizebox{\hsize}{!}{\includegraphics{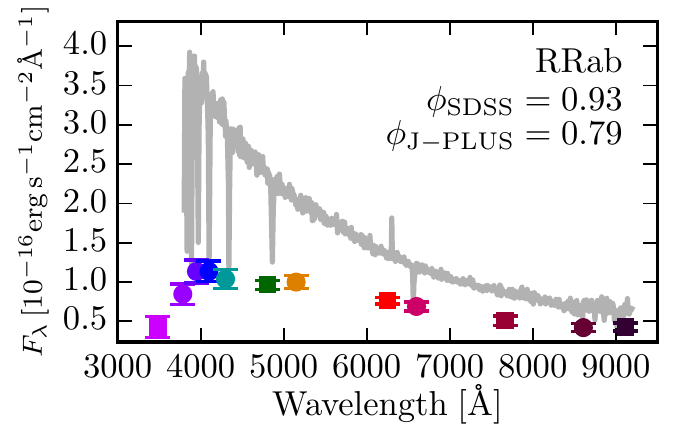}}\\ \resizebox{\hsize}{!}{\includegraphics{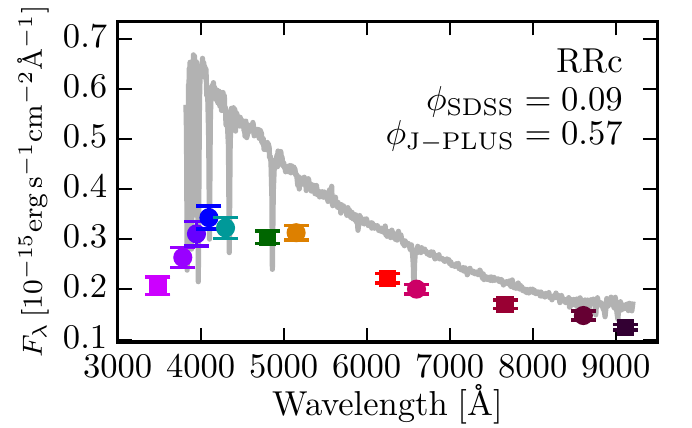}} \caption{\jplus 	photo-spectra of two RR Lyrae stars, one type RRAb ({\it top
	panel}; $\alpha = 122.9568$\,deg, $\delta = 31.089$\,deg) and
	one type RRc ({\it bottom panel}; $\alpha= 123.6246$\,deg,
	$\delta= 31.188$\,deg). The gray lines show the SDSS spectra
	of these stars. The \jplus photo-spectra and the SDSS spectra
	were acquired at different pulsation phases $\phi$, as labeled
	in the panels, with $\phi = 0,\,1$ corresponding to the
	maximum flux of the source (i.e., the entire period is
	covered by the $\phi$ range of values $\phi \in [0,1)$. Both
	SDSS spectra were taken not far from their respective maximum
	brightnesses.}  \label{fig:rrlyrae}
\end{figure}

On the observational side, their pulsation amplitude (from 0.2 to 1.5 magnitudes)
and period (from 5\,h to 1.2\,days) make their variability relatively easy
to be detected using both, \jplus and external data archives. Additionally,
\jplus will provide the SED of a unprecedented amount of RR Lyrae stars. 
Two examples are shown in Fig.~\ref{fig:rrlyrae}, taken from the \jplus EDR.

\section{Summary and conclusions}
\label{sec:conclusions}

This paper is devoted to present and illustrate the \jplus survey and its main scientific applications. \jplus is a 12-band photometric survey of thousands of square degrees in the Northern hemisphere which is being conducted at the OAJ. The most remarkable characteristics of the survey rely on the 2\,deg$^2$ FoV of T80Cam at the JAST/T80 telescope and on a unique system of 12 optical filters, including broad band, medium width, and narrow band optical filters in the range $3\,500$--$10\,000$\,\AA. This filter system has been designed to optimally extract rest-frame spectral features that are key to characterize the stellar types in our MW as well as to perform stellar population and star formation studies in nearby galaxies. Overall, \jplus provides an unprecedented low resolution photo-spectrum of 12 points for every pixel of the sky in the footprint of the survey. 

The paper is also aimed to release the first 36\,deg$^2$ of \jplus data, amounting to $\sim 1.5\times 10^5$ stars and $10^5$ galaxies at $r<21$\,mag observed in 12 broad, intermediate and narrow optical bands. This dataset is used to illustrate some of the science cases that will be addressed by \jplus. As of early 2018, \jplus has already mapped more than $1\,000$\,deg$^2$, providing unprecedented information of the SED for millions of stars and galaxies. A forthcoming data release with a significantly larger area to the one presented in this paper is expected to be published along 2018.

With a choice of narrow and medium width filters strategically placed to render accurate stellar type information (the 3\,700--4\,000\,\AA\, Balmer break region, H$\delta$, Ca H+K, the G-band, the Mg$b$ and Ca triplets), the addition of narrow band filters centred upon the H$\alpha$/$\lambda$6563 and [OII]/$\lambda$3727 lines constitutes a powerful window to the star formation activity in the local universe, and also to bright emission lines at higher redshifts. Consequently, \jplus is providing a unique data set with direct applications in the characterization of the stellar populations of our Galaxy, including metal poor stars, white dwarfs, ultracool dwarfs, planetary nebulae, symbiotic stars, cataclismic variables, and globular clusters. The same set of filters enable IFU-like 2D studies in nearby ($z<0.015$) galaxies with unprecedented quality and statistics, and provides high-quality photo-$z$ estimates (with an error amplitude of $\delta z \sim 0.01$--$0.03$ for $r<21$\,mag) for extragalactic objects. In addition, in the extragalactic sky, \jplus' narrow band filters provide sensitivity to particular redshift shells in the universe corresponding to the rest-frame frequency of power lines like Ly$\alpha$ or [OII]/$\lambda$3727.  

A great part of the added value of \jplus resides, besides its large scale/panoramic character owed to its 2\,deg$^2$ FoV, in the medium-width and narrow band filters. These bring \jplus along the way of other spectro-photometric surveys where multiple band imaging is used as an alternative to spectroscopy. The indiscriminate nature of this approach, which drastically simplifies selection biases, together with the significantly larger statistics of the resulting catalogues, constitute two powerful arguments supporting this strategy. While the upcoming \jpas project can be seen as the definite test where all these expectations will be confronted with real data,  \jplus is currently paving exactly the same way but under more modest instrumentation. Nonetheless, despite the intrinsic scientific interest of \jplus, it turns out to be an ideal, strategical test bench for \jpas as most technical and software developments of \jplus will apply very similarly to \jpas. In addition, all these efforts are being complemented by the twin project S-PLUS, which, from Cerro Tololo, is scanning the Southern hemisphere with a replica of the \jplus telescope, camera and filter system.

The combination of this type of surveys with spectroscopic ones for calibration and comparison purposes, and from other surveys at other wavelength ranges such as X-ray or $\gamma$-ray, should significantly improve our knowledge of the SEDs for practically {\em all} families of sources in our universe.  Furthermore, if systematics are kept under the required level, this spectro-photometric approach should also open the possibility to study the unresolved emission associated to particular emission lines at particular redshift shells \citep[aka intensity mapping,][]{kovetz2017}, thus providing a new way to conduct tomography of the universe in the optical range. 

To conclude, \jplus may thus be opening an exciting and interesting new phase in optical, large scale, astrophysical surveys. Ultimately, \jplus will become a powerful multicolor view of the nearby Universe that will observe and characterize tens of millions of galaxies and stars of the MW halo, with a wide range of astrophysical applications and a striking potential for bringing unexpected discoveries to our knowledge of the Universe. The \jplus data will be made public progressively in subsequent data releases, an expression of its commitment to become a major legacy project for the astronomy and astrophysics of the next decades. 

\begin{acknowledgements}
Funding for the \jplus Project has been provided by the Governments of Spain and Arag\'on through the Fondo de Inversiones de Teruel, the Spanish Ministry of Economy and Competitiveness (MINECO; under grants AYA2017-86274-P, AYA2016-77846-P, AYA2016-77237-C3-1-P, AYA2015-66211-C2-1-P, AYA2015-66211-C2-2, AYA2012-30789, AGAUR grant SGR-661/2017, and ICTS-2009-14), and European FEDER funding (FCDD10-4E-867, FCDD13-4E-2685). Further support has been provided by the Ram\'on y Cajal programmes  RYC-2016-20254, RYC-2011-08262 and RYC-2011-08529.

This research has made use of the Spanish Virtual Observatory (\url{http://svo.cab.inta-csic.es}) supported from the Spanish MINECO through grant AYA2014-55216. We also acknowledge Spanish CSIC (I-COOP+ 2016 program) through grant COOPB20263. 
The Brazilian agencies FAPESP and the National Observatory of Brazil have also contributed to this project. We acknowledge financial support from the Funda\c c\~ao Carlos Chagas Filho de Amparo \`a Pesquisa do Estado do Rio de Janeiro - FAPERJ (fellowship Nota 10, PDR-10), from CNPq through BP grant  312307/2015-2 and Universal Grants 459553/2014-3, PQ 302037/2015-2, and PDE 200289/2017-9), FINEP grants  REF. 1217/13 - 01.13.0279.00 and REF 0859/10 - 01.10.0663.00, from FAPERJ grant E-26/202.835/2016, and  
CAPES (Science without Borders program, Young Talent Fellowship, BJT) through grants A062/2013 and CAPES-PNPD 2940/2011. The FAPESP grants no. 2015/12745-6, 2014/11338-5, 2014/07684-5, 2013/04582-4 and 2009/54202-8 are also acknowledged. Finally, the authors acknowledge partial support from grant PHY 14-30152; Physics Frontier Center/JINA Center for the Evolution of the Elements (JINA-CEE), awarded by the US National Science Foundation.

\end{acknowledgements}

\bibliographystyle{aa}
\bibliography{ms}
\end{document}